\documentclass[final,5p,times,twocolumn]{elsarticle}

\usepackage{amssymb}

\usepackage{amsmath,amssymb,amsfonts}
\usepackage{algorithmic}
\usepackage{graphicx}
\usepackage{textcomp}
\usepackage{booktabs}
\usepackage{hyperref,comment}   
\usepackage{tabularray}
\usepackage{algorithm}
\usepackage{ulem}
\usepackage{float}
\usepackage{array}
\usepackage{colortbl}
\usepackage[table,xcdraw]{xcolor}

\usepackage{caption}

\def \myver{2}

\def \hugever{3}

\newcommand{\ar}[1]{\textcolor{black}{#1}}

\newcommand{\hw}[1]{\textcolor{black}{#1}}

\newcommand{\arr}[1]{\textcolor{red}{\textit{Razi: #1}}}

\newcommand{\fullyr}[1]{\textcolor{red}{Razi: This section is fully revised. don't change!}}
\newcommand{\uhere}[1]{\textcolor{red}{Razi: Razi revised until here!}}
\newcommand{\underr}[1]{\textcolor{red}{Razi: This section is under revision. change only with slash hw! dont remove any considerable text and use slash sout if you want to remove}}

\journal{Applied Energy}

\begin{document}

\begin{frontmatter}

\title{Energy Optimization for HVAC Systems in Multi-VAV Open Offices: A Deep Reinforcement Learning Approach}

\author[inst1]{Hao Wang}
\author[inst1]{Xiwen Chen}

\affiliation[inst1]{organization={School of Computing, Clemson University},
            addressline={821 McMillan Rd}, 
            city={Clemson},
            postcode={29631}, 
            state={SC},
            country={USA}}

\author[inst2]{Natan Vital}
\author[inst2]{Edward Duffy}
\author[inst1]{Abolfazl Razi}

\affiliation[inst2]{organization={BMW Information Technology Research Center, CU-ICAR},
            addressline={2 Research Dr}, 
            city={Greenville},
            postcode={29607}, 
            state={SC},
            country={USA}}

\begin{abstract}

With global warming intensifying and resource conflicts escalating, the world is undergoing a transformative shift toward sustainable practices and energy-efficient solutions. With more than 32\% of the global energy used by commercial and residential buildings, there is an urgent need to revisit traditional approaches to Building Energy Management (BEM). Within a BEMSplatform, regulating the operation of Heating, Ventilation, and Air Conditioning (HVAC) systems is more important, noting that HVAC systems account for about 40\% of the total energy cost in the commercial sector. 

This paper offers a Deep Reinforcement Learning (DRL) algorithm as a data-driven approach to controlling HVAC operation to enhance the energy efficiency of commercial buildings with open offices while ensuring thermal comfort for occupants in different zones. 
Compared to alternative methods such as rule-based models and model-predictive control, data-driven models have shown promising results in optimizing building energy consumption without the need for building-specific thresholds, prior knowledge about the underlying physics of heat distribution, and digital mapping of the airflow. 
Despite the astonishing performance of modern DRL methods in controlling energy management, a particular energy-saving solution for open-plan offices with multiple Variable Air Volume (VAV) systems, where different zones can not be treated independently, is still missing. Also, some of the existing methods suffer from key issues such as long training time and lack of generalizability for using over-complicated models, incorporating external factors that are hard to model and characterize, and including factors that are not typically accessible. 
To solve these issues, we propose a low-complexity DRL-based model with multi-input multi-output architecture for the HVAC energy optimization of open-plan offices, which uses only a handful of controllable and accessible factors. 
The efficacy of our solution is evaluated through extensive analysis of the overall energy consumption and thermal comfort levels compared to a baseline system based on the existing HVAC schedule \hw{from a real case}.
This comparison shows that our method achieves 37\% savings in energy consumption with minimum \hw{temperature} violation ($<$1\%) of the desired temperature range during work hours. It takes only a total of 40 minutes for 5 epochs (about 7.75 minutes per epoch) to train a network with superior performance and covering diverse conditions for its low-complexity architecture; therefore, it easily adapts to changes in the building setups, weather conditions, occupancy rate, etc. 
Moreover, by enforcing smoothness on the control strategy, we suppress the frequent and unpleasant on/off transitions on HVAC units to avoid occupant discomfort and potential damage to the system.
The generalizability of our model is verified by applying it to different building models and under various weather conditions.


\end{abstract}

\begin{graphicalabstract}
\end{graphicalabstract}

\begin{highlights}

\item Model and analyze the thermodynamics of multi-zone open-plan offices. 
\item Present a DRL-based control algorithm that simultaneously optimizes thermal comfort and energy efficiency. 
\item Achieve a 37\% reduction in HVAC energy with less than 1\% temperature violation. 
\item Propose a heuristic reward policy and smooth control \hw{algorithm} to minimize on-off transitions. 
\item Validate the generalizability of the proposed method to a different floor plan under various weather conditions.

\end{highlights}


\end{frontmatter}








\section{Introduction}
\label{sec:introduction}



\begin{figure*}[htbp]
    \centering
    \centerline{\includegraphics[width=1\textwidth]{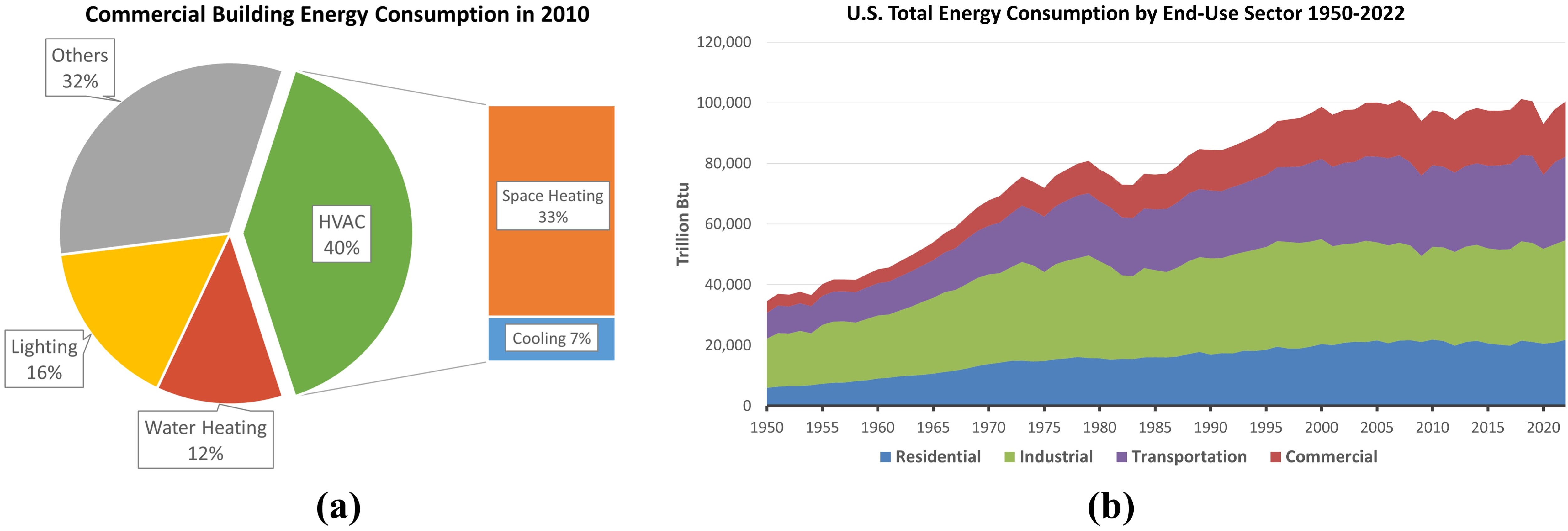}}
    \caption{(a) Final building energy consumption in the world by end-use in 2010 \cite{urge2015heating}, and (b) the trend of energy used by different end-use sectors in the USA during the last 70 years\cite{eia2022energy}.}
    \label{fig:facility_E}
\end{figure*}


The global energy outlook is witnessing a continuous surge in the demand for energy resources, alongside a growing concern for environmental sustainability. With the world's population reaching 8 billion people \cite{population2022}, our consumption of energy has reached unprecedented levels. The excessive use of energy sources like coal, oil, and natural gas has resulted in the release of harmful greenhouse gases, contributing significantly to the critical issue of global warming. These emissions not only deplete our planet's limited resources but also exacerbate conflicts related to resource availability. This highlights the importance of more informed energy consumption, worldwide.

\subsection{Energy Consumption in Commercial Buildings}

Among all sectors, residential and commercial buildings are key areas for energy conservation since they account for a substantial portion of the world's energy usage.  According to the global energy statistics up to 2010, the building sector accounted for roughly one-third of the total energy demand, with 24\% attributed to residential buildings and 8\% to commercial buildings \cite{urge2015heating}. 
By 2021, the energy consumption of commercial buildings had risen from 8\% to 18\% \cite{eia2022energy}, highlighting the importance of energy management for commercial buildings. Figure \ref{fig:facility_E} (b) illustrates the trend of energy use in end-use sectors throughout the United States.

Within a building, the Heating, Ventilation, and Air Conditioning (HVAC) system is one of the most vital components used to circulate air, filter dust, and retain thermal comfort for the occupants by controlling the temperature and oxygen levels, in different zones of the building. 
The air ventilation capabilities of HVAC systems are even more important in the industry and commercial sector 
to construct clean rooms such as medical facilities, optical laboratories, and bio-hazard-related rooms. Inadequate ventilation or inappropriate use of HVAC systems may raise severe health issues and even harm the safety of occupants. For instance, a recent study has shown that elevated levels of CO2 levels are correlated with short-term sickness in office spaces \cite{azuma2018effects}.

However, the significance of HVAC systems extends beyond indoor air quality and thermal comfort. HVAC systems account for up to approximately 40\% of the total energy usage in commercial buildings \cite{urge2015heating}, as shown in Figure \ref{fig:facility_E} (a). Consequently, optimizing HVAC energy, as studied in this paper, is imperative in achieving energy efficiency goals and promoting sustainable practices in commercial building operations.

\begin{table*}
\centering
\caption{\large Summary of Building Energy Optimization Approaches }
\label{tab:review}
\resizebox{0.8\linewidth}{!}{%
\begin{tblr}{
  width = \linewidth,
  colspec = {Q[120]Q[280]Q[150]Q[100]Q[70]},
  cells = {c},
  cell{2}{1} = {r=3}{},
  cell{2}{2} = {r=3}{},
  cell{5}{1} = {r=5}{},
  cell{5}{2} = {r=5}{},
  cell{6}{3} = {r=2}{},
  cell{10}{1} = {r=7}{},
  cell{10}{2} = {r=7}{},
  cell{12}{3} = {r=5}{},
  vlines,
  hline{1,17} = {-}{0.08em},
  hline{2,5,10} = {-}{},
  hline{3-4,6,8-9,11-12} = {3-5}{},
  hline{7,13-16} = {4-5}{},
}
Category                       & Features                                                                                                                                                                                                                                                & Tasks                            & Approach        & Reference                 \\
Rule-Based Control (RBC)       & Traditional approaches to optimize  building energy consumption by monitoring, controlling, and regulating its systems and equipment based on predetermined rules or logic                                                                              & Lighting control                 & Grey Prediction & \cite{leephakpreeda2005adaptive} \\
                               &                                                                                                                                                                                                                                                         & HVAC Control~ ~                  & PID             & \cite{homod2009pid}              \\
                               &                                                                                                                                                                                                                                                         & PV Control                       & RBC             & \cite{salpakari2016optimal}      \\
Model Predictive Control (MPC) & Mathematical models that predict future behavior and optimize control inputs to achieve desired energy efficiency in real-time.                                                                                                                         & Supply Water Temperature Control & PSO             & \cite{corbin2013model}           \\
                               &                                                                                                                                                                                                                                                         & HVAC Control                     & NLMPC           & \cite{mantovani2014temperature}  \\
                               &                                                                                                                                                                                                                                                         &                                  & ADMM            & \cite{hou2016distributed}        \\
                               &                                                                                                                                                                                                                                                         & Building Control                 & MOPSO           & \cite{delgarm2016multi}          \\
                               &                                                                                                                                                                                                                                                         & Temperature/Air Flow Prediction   & RL-MPC          & \cite{arroyo2022reinforced}      \\
Reinforcement Learning (RL)    & Using Deep Learning (DL) and/or Reinforcement Learning (RL) methods to optimize energy consumption by data-driven modeling of building systems that implement automatic and continuously improving control strategies based on action-reward mechanisms & Building Control                 & DPG             & \cite{mocanu2018line}            \\
                               &                                                                                                                                                                                                                                                         & PV Control                       & FH-DDPG         & \cite{lei2020dynamic}            \\
                               &                                                                                                                                                                                                                                                         & HVAC Control                     & DQN             & \cite{wei2017deep}               \\
                               &                                                                                                                                                                                                                                                         &                                  & A3C             & \cite{zhang2018practical}        \\
                               &                                                                                                                                                                                                                                                         &                                  & DDPG            & \cite{du2021intelligent}         \\
                               &                                                                                                                                                                                                                                                         &                                  & LSTM-DDPG       & \cite{zou2020towards}            \\
                               &                                                                                                                                                                                                                                                         &                                  & MA-DRL          & \cite{yu2020multi}               
\end{tblr}}
\end{table*}

\subsection{Post-COVID Requirements}
\label{sec:covid}


Achieving significant reductions in the energy cost of HVAC systems is a challenging task. HVAC energy management has become even more imperative in the post-Covid era since a lot of companies have adopted remote working policies to prevent the spread of infection \cite{cai2022impact}. As a result, daily occupancy in offices has reduced to half or even less \cite{su132111586}. However, an office building still needs to spend at least 20\% of its energy to maintain its fundamental functionality even when it is unoccupied \cite{su132111586}. Despite the drastic decrease in occupancy rates, energy consumption in commercial buildings has not shown a significant decline as HVAC systems still run at the same pace regardless of the occupancy rates. Therefore, it is imperative to develop optimal operation control for HVAC systems that balance energy efficiency and thermal comfort for the remaining occupants. 

\subsection{Building Energy Management Systems}
\label{sec:BEMS}

To optimize energy consumption in commercial buildings, Building Energy Management Systems (BEMS) have been developed. BEMS integrates various technologies, such as sensors, data analysis tools, and control algorithms, to monitor, analyze, and control energy-consuming systems. Contemporary commercial buildings equipped with BEMS can make use of smart sensors to dynamically adjust energy consumption based on the occupancy rate and other factors. Furthermore, the centralization of HVAC systems in these buildings allows for the implementation of more sophisticated algorithms. Overall, the BEMS provides valuable insights into energy usage patterns, identify energy-saving opportunities, and enables proactive energy management strategies.

Generally, there exist the following three different mainstream approaches for energy control in BEMS shown also in Table \ref{tab:review}.

\textbf{A) BEMS with Rule-Based Control:} 
\label{sec:RBC}
The conventional approach to building control is using Rule-Based Control (RBC) strategies. This approach is favored by many people for its simplicity and reliability. These policies are often designed by experts and operational forces using empirical data and engineering experiences \cite{ASHRAE202262}. 
For instance, \cite{leephakpreeda2005adaptive} applied an adaptive occupancy-based lighting control with a grey prediction model. On the other hand, a cascade PID controller is proposed to control the HVAC system in \cite{homod2009pid}.

However, as commercial building complexity continues to increase, the inflexibility of these rule-based strategies can result in lower energy efficiency. 
As a result, a considerable amount of energy is still being wasted through various means such as the inadequate optimization of unoccupied spaces, the preservation of thermal comfort during non-working hours, and the adoption of inappropriate policies in functionally-deficient areas such as restrooms and storage facilities. Therefore, inefficient control policies and invalid thermal retention practices seem to be the primary contributors to energy waste. 
Thus, achieving a balance between optimizing energy usage and ensuring occupants' health and comfort is a challenging task that is not best addressed by rule-based control systems.

\textbf{B) Model Predictive Control-based BEMS:}
\label{sec:MPC}
An alternative energy management methodology is using Model Predictive Control (MPC), which uses the learning power of Machine Learning (ML) algorithms to predict potential outcomes of the energy management systems. MPC has made significant progress in recent years, leading to the development of sophisticated HVAC control policies and algorithms that have shown promising results in energy optimization \hw{ \cite{corbin2013model, mantovani2014temperature, hou2016distributed, yao2021state}}. The MPC approach involves creating a digital twin of the real building, which closely mirrors the building structure and control logic. This digital model enables the prediction of temperature dynamics and facilitates the analysis of thermal behavior. By utilizing the digital twin model, MPC optimizes the control inputs, such as setpoints and actuator actions, to achieve the desired objectives of energy efficiency and occupant comfort, while considering operational constraints. 

Meanwhile, optimization-based building control, as a sub-MPC method, applies meta-heuristic optimization algorithms such as Particle Swarm Optimization (PSO) \cite{delgarm2016multi, mehrabi2021age} to find the optimal control inputs within the constraints defined by the MPC framework. 
However, despite the accuracy of MPC-based building energy control, HVAC systems operate under dynamic conditions influenced by various external factors such as outdoor temperature, solar radiation, occupancy patterns, and internal heat gains, which would be hard to include in MPC, due to their complex underlying physics. 
Additionally, constructing and maintaining a valid and accurate digital model of a complex commercial building can be a tedious and time-consuming task. The last issue is the lack of generalizability since we need to build a separate model for each building.

\textbf{C) Reinforcement Learning-based BEMS:} 
\label{sec:RL}
The third mainstream approach in the field of energy management is using data-driven Reinforcement Learning (RL) techniques to control energy suppliers and users in complex environments. RL methods enable the agents (e.g., HVAC controllers in our case) to learn optimal policies by observing their interaction with the environment (e.g., the temperature map of the target zone in our case). One of the key advantages of RL methods is that they do not require prior knowledge about the complex physics of heat conduction models or external factors, making them flexible and adaptable to various settings. As a result, data-driven approaches have recently gained significant attention from the research community \cite{wei2017deep, wei2019deep, fang2022deep, yu2021review, fu2022applications}. 

Due to the success of DRL, it has become widely adopted in energy management applications. 
For instance, a Deep Q-Network (DQN) with a memory buffer is implemented in \cite{wei2017deep} to control the airflow rate in different zones. Their results demonstrate the utility of DRL as an efficient helper in solving energy optimization problems in complex environments. 
Zhang et al. applied an improved DRL algorithm, called Asynchronous Advantage Actor Critic (A3C) to an actual building to control the supply water temperature\cite{zhang2018practical}. Combined with the model prediction control, their proposed method achieved a significant reduction in energy consumption while maintaining thermal comfort.
Likewise, Deep Deterministic Policy Gradient (DDPG) is applied to an HVAC system to address the limitation of DQN in continuous control policies \cite{du2021intelligent}. Additionally, \cite{yu2020multi} utilized Multi-Agent Deep Reinforcement Learning (MADRL) to optimize the building energy optimization by controlling the supply air rate and damper position. Their proposed method approached a considerable energy saving in a 30-zone commercial building. 


\subsection{Current Challenges}
\label{sec:challenge}

Despite the promising results of DRL methods to optimize energy consumption in buildings, the majority of prior DRL methods are implemented only in simulation environments (e.g., \cite{wei2017deep, mocanu2018line, yoon2019performance, zou2020towards, du2021intelligent,deng2022towards}); therefore, they are not practically efficient due to following drawbacks \cite{yu2021review}:

One common problem of the current DRL methods is the excessive use of impractical variables, which creates challenges for practical implementation.
For instance, \cite{deng2022towards} utilized nine environmental variables (such as diffuse solar radiation and wind direction) and six system states for a 5-zone building. Obtaining such data in real-time can be difficult in practice. 
Additionally, the method proposed in \cite{yu2020multi} relies on the number of occupants, which might be hard to trace accurately due to the random patterns in occupancy and lack of precise occupant tracking sensors in most buildings. They also incorporate variables such as electricity prices which may vary across different geographical regions that prevent the generalizability of the model.

Deploying over-complicated DRL models with a huge number of network parameters can improve the results but at the cost of prohibitively long training time. 
For instance, the model proposed in \cite{zhang2018practical} takes 10 hours for complete training, and the model in \cite{du2021intelligent} takes 50 episodes to converge. 
This extended training duration restricts the applicability of such methods in real-world scenarios where time-sensitive decision-making is required, or the model needs to adapt to frequent changes in the environment. 
We implement a DQN model with only 14 input factors (2 global factors and 2 per-zone factors for 6 zones) and with approximately 18,818 parameters for the utilized neural network. Our network takes only 40 minutes to converge using a typical computer. Also, our model is readily adaptable to different building scenarios, including open-floor plans.

Additionally, BEMS may restrict access to control variables, and regulations such as ASHRAE standards limit the manipulation of supply water temperature and ventilation dampers to prevent thermal discomfort for occupants \cite{ASHRAE202262, ASHRAE201755}. In contrast to prior work, we exclude such control variables from the optimization process since manipulating any restricted factors such as the air flow rate of VAV units or the chill water temperature of the Air Handling Unit (AHU) may raise severe safety issues \cite{azuma2018effects}.

\subsection{Research Gap in Open-plan Offices}
\label{sec:gap}

\begin{figure}[htbp]
    \centering
    \centerline{\includegraphics[width=1\columnwidth]{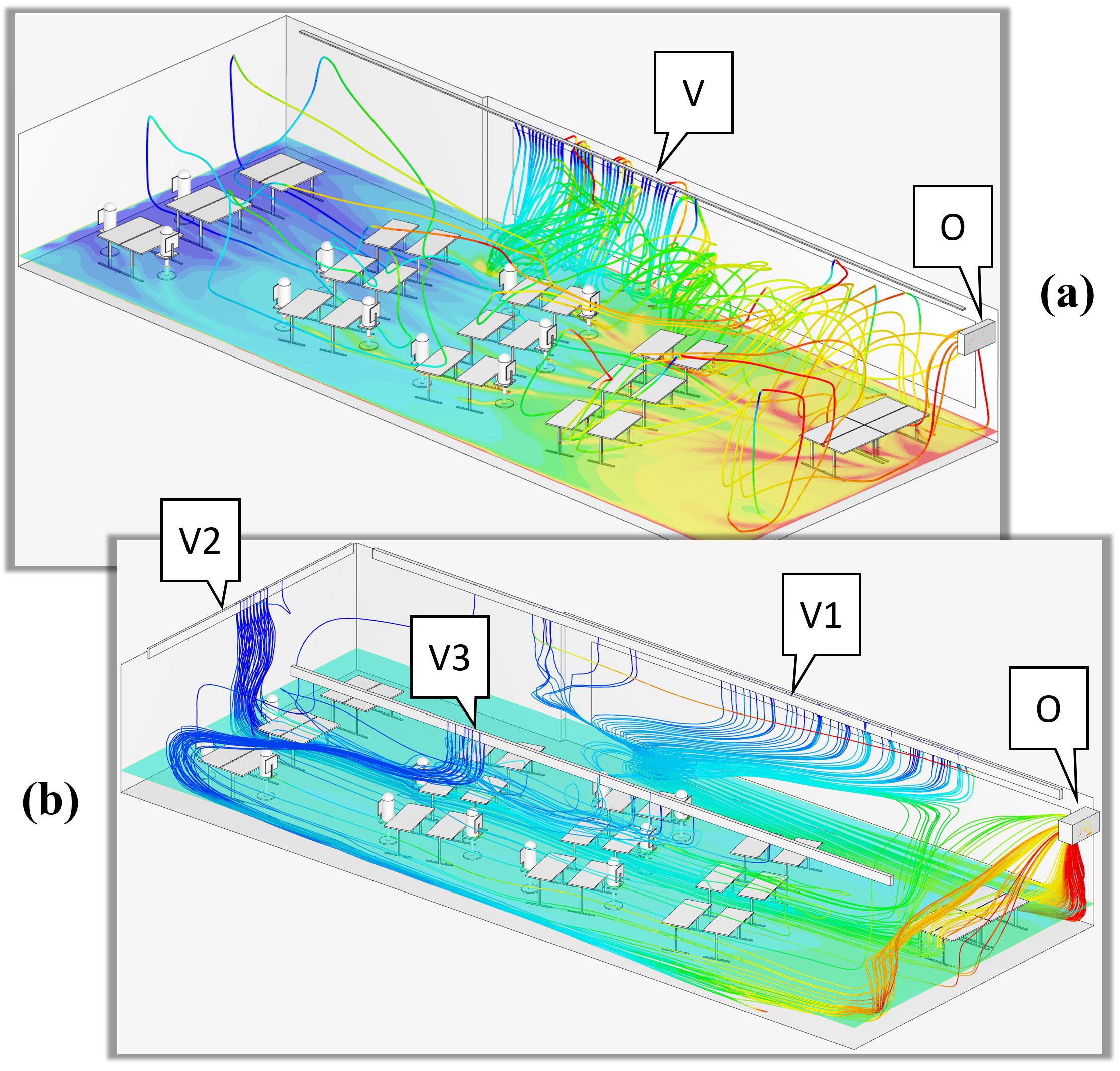}}
    \caption{\ar{Open Office Model in 3D view. (a) Open Office with single Linear Slot Diffuser VAV, (b) Open Office with Multi-VAV Systems. V and O represent the vent and the outlet.} }
    \label{fig:office}
\end{figure}

    

Open-plan offices, where the interior walls are moved to create a large communal workspace, have gained popularity in modern commercial buildings \cite{lai2021open}. Research has also shown that the open-plan design in offices can lead to increased productivity and communication efficiency \cite{shafaghat2014open}. 
Such configurations are often equipped with multiple Variable Airflow Volume (VAV) units to regulate the temperature in multiple zones to achieve better heat transfer, as a significant factor in reducing the building's overall energy consumption \cite{bodart2002global}. Figure \ref{fig:office} shows the high heat transfer efficiency of the multi-VAV system in open offices compared to the single-VAV system. 

Despite the popularity of open-plan offices in commercial buildings, limited research has been conducted to address the importance of energy optimization in these types of spaces. For instance, VAV units in such offices often operate independently, without considering the interconnectivity of these spaces, which can result in a disparity in heating and cooling, with areas located close to vents receiving more ventilation-based heating/cooling, while spaces near windows receive more heat from solar radiation.

In this study, we present a DRL-based HVAC control method to optimize building energy consumption in such floor plans. Our specifically designed open office model consists of multiple interconnected spaces, and the DRL algorithm is applied to control multiple VAV units jointly.

\begin{figure*}[htbp]
    \centering
    \centerline{\includegraphics[width=1\textwidth]{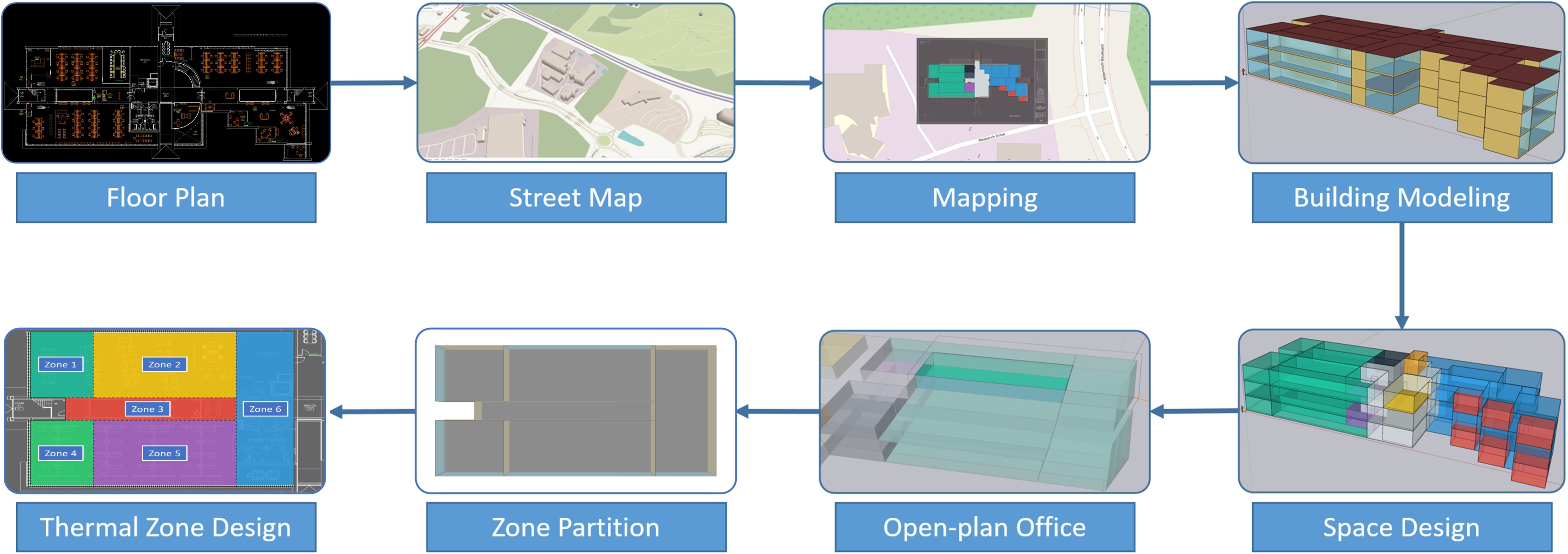}}
    \caption{Framework of building modeling.}
    \label{fig:building}
\end{figure*}

\subsection{Summary of Contributions}  
\label{sec:summary}
In short, the contributions of this paper can be summarized as follows:

\begin{itemize}

\item We analyze the heat transfer features of connected spaces in open-plan offices and compare their energy consumption to offices with traditional closed floor plans. \ar{We offer a formulation for thermal energy exchange that suits open offices.}

\item We propose a DRL-based control algorithm that simultaneously optimizes thermal comfort and energy efficiency using a multiple-input and multiple-output architecture. It resulted in a 37\% reduction in HVAC energy consumption with less than 1\% violation of the temperature comfort level \ar{and 2.5\% violation of humidity comfort level. Note that our model is flexible and can trade off energy efficiency with comfort violation by controlling the tuning parameters}.  

\item The proposed model requires only minimal input variables, including the outdoor temperature, indoor temperature, time, and control signals. \ar{The action space is a binary vector to activate/inactivate enforcing temperature range, instead of using explicit set points. These two approaches make the framework concise and easily} generalizable to other buildings. 

\item We apply a heuristic reward policy to accelerate the training process and reduce the model complexity.

\item We introduce a penalty term in the cost function that penalizes frequent inconvenient on/off transitions to avoid discomfort and damage to the HVAC system. 



\item Our model is computationally efficient and takes only about 7.75 minutes per epoch (about 40 minutes for 5 epochs) to train. It can be easily adapted to other open-plan offices, making it a universal solution for building energy optimization.

\end{itemize}

The remainder of this paper is organized as follows: In Section \ref{sec:work}, we provide a detailed explanation of the utilized building model. 
Then, we review DRL methods and show how Markov Decision Processes (MDP) can be used to model temperature variation by HVAC status change.
In Section \ref{sec:problem}, we state our hypothesis that open offices have better heat transfer performance. 
In Section \ref{sec:solution}, we provide the details of our solution to the open office simulation, HVAC control strategies, and DRL model design. 
In Section \ref{sec:case}, we present a case study and evaluate the effectiveness of the proposed DRL model, examining the energy characteristics of an open office in comparison to conventional office designs. Finally, in Section \ref{sec:results}, we discuss the results and analyze the impact of temperature and signal smoothness trade-off on energy optimization, and discuss how the DRL model minimizes energy consumption.

\section{Background Information} 
\label{sec:work}

Modeling involves creating mathematical models that represent the internal thermodynamics and interactions between the building systems, such as HVAC and lighting. It also includes characterizing energy consumption and other performance metrics.



\subsection{Model-based Simulation}


Some recent works have demonstrated the potential of information-based neural networks in Computational Fluid Dynamics (CFD) and Finite Element Analysis (FEA) \cite{cai2021physics, liang2018deep}, yielding increased accuracy and faster computation. 
However, model-based predictions tend to focus on the features of materials and physical models within a specific environment. Energy consumption in buildings, on the other hand, is influenced by various external factors, making it difficult to accurately predict through simulations. For instance, Mantovani et al. proposed a digital building model to simulate energy consumption \cite{mantovani2014temperature}. Despite being highly consistent with the real building in terms of the floor plan, layout, and zone design, the accuracy of their simulation was impacted by changes in weather, season, and HVAC system conditions. The heat transfer model in the real world is more complex and cannot be fully replicated, leading to an increased error.

In this paper, we address the limitations of traditional building energy modeling and offer a simplified approach by focusing on the heat transfer characteristics in open-plan offices instead of developing a highly detailed building model. 
Specifically, our modeling is based on the hypothesis that open floor plans allow greater heat exchange between adjacent zones through airflow compared to closed offices that offer better heat isolation with solid walls. 
To validate this hypothesis, we modeled \textbf{BMW Information Technology Research Center} (BMW-ITRC), a contemporary commercial building in Greenville, SC that is equipped with smart sensors and a modern BEMS developed by ICONICS. The building model is based on the actual geographic location and historical weather data of the local city. We map the floor plan to the Open Street Map (OSM) to ensure the size of the building model is close to the actual building.
Additionally, we build a simplified open office model with a spatial design that is consistent with the building in the real world. Figure \ref{fig:building} shows the workflow of the open office modeling.

\begin{figure}[htbp]
    \centering
    \centerline{\includegraphics[width=1\columnwidth]{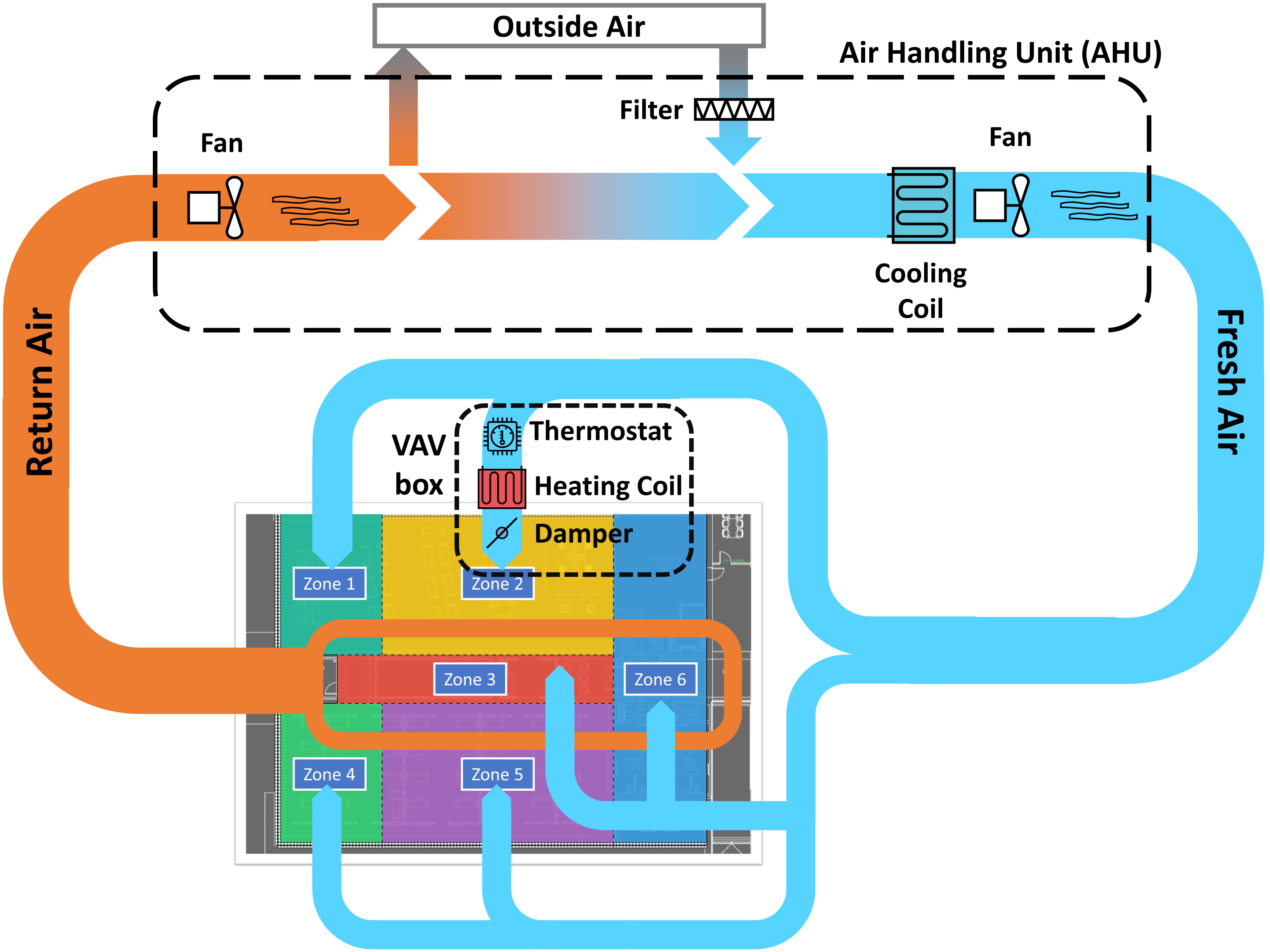}}
    \caption{HVAC system with multi-VAV units in commercial buildings.}
    \label{fig:VAV}
\end{figure}

\subsection{HVAC Operation}
\label{sec:hvac}

HVAC systems function by effectively regulating the indoor environment, ensuring desired temperature and air quality for occupants. The Air Handling Unit (AHU) is responsible for bringing in outside air and conditioning it. The AHU typically includes filters to remove pollutants and dust, ensuring the air quality meets the required standards. The conditioned air is then distributed to different zones within the building through a network of ducts. Each zone is equipped with Variable Air Volume (VAV) units, which control the amount of airflow and the temperature of the supplied air based on zone-specific requirements. The VAV units may include reheat coils that adjust the temperature of the air to achieve optimal comfort \hw{ \cite{okochi2016review}}. 
By delivering conditioned air to individual zones, the HVAC system ensures that each area can be maintained at the desired temperature while providing fresh and healthy air to occupants. Figure \ref{fig:VAV} shows a typical design of multi-VAV systems \cite{LEGG201753, ASHRAE202262}.

\subsection{Reinforcement Learning} \label{sec:rl}

Reinforcement Learning is a tool that can be used to obtain optimal policies for problems that follow probabilistic state transitions, usually modeled by a Markov Decision Process (MDP). 
RL is appropriate when the action-reward relations are not priory known but can be inferred by observing episodes of state-action-rewards. For instance, in our problem, the state of the system at time $t$, $S_t$ is a vector that represents the temperature and current HVAC status of different zones. In our case, it also includes some uncontrollable system input like the outdoor temperature and work hour indicator. 
Likewise, the action at time $t$, $a_t$ is the control signal, which in our case is turning on and off the comfort policy to different VAV units, which further translates to set-points and physical tasks like turning on and off or changing duty cycle of the heater and cooler systems. The goal of RL is finding an optimal policy $\pi:\mathcal{S} \mapsto \mathcal{A}$ that maps state $s \in \mathcal{S}$ to action $a \in \mathcal{A}$, so that the accumulated reward for starting from some initial state $s_0$ and following policy $\pi$ (taking actions as $a_t=\pi(s_t)$) until a finite or infinite horizon $T$ is maximized. In other words, we aim to maximize
\begin{align}
\pi* = \underset{\pi}{agrmax} ~ \text{E} \sum_{t-0}^T \gamma^k R(s_t,a_t=\pi(s_t),s_{t+1}),
\end{align} 
where $\gamma$ is the discount factor to promote faster results and $E[]$ is the expected value noting that transitions under MDP are probabilistic and defined by transition probabilities $P(s_{t+1}|s_t,a_t)$. In energy management, heat transfer equations, air flow simulators, observing historical data of a real system, or a combination of them are used to model transitions. In our case, we use EnergyPlus simulator along with the floor plan and HVAC operation schedule. 
The reward $r_t=R(s_t,a_t,s_{t+1})$ represents the obtained reward at time $t$ which reflects the desirability of transitioning from state $s_t$ to state $s_{t+1}$ by taking action $a_t$. In some cases, including ours, we can simply use it to evaluate the desirability of the current state as $r_t=R(s_t)$, for example, it can be a numerical value that quantifies the compliance of the temperature with desired limits while considering the energy cost. We design a heuristic reward to enforce different objectives as detailed in Section \ref{sec:reward}.

In this work, we use deep Q-learning to solve MDP. Q-learning is one of the most widely used RL methods, where the quality of action $a$ in a given state $s$ is captured by the Q-value $Q(s,a)$. The Q-values are updated based on the sequences of (state, action, reward) tuples observed during the operation or simulation-based training phase. The Q-value update follows the equation $Q_{t+1}(s,a)=(1-\alpha) Q_{t}(s,a) + \alpha \left(r + \gamma \max_{a'} Q_t(s',a')\right)$, where $\alpha,~ 0<\alpha<1$ is the learning rate. Q-learning is appreciated for its simplicity, although it faces challenges when the number of states becomes extremely large, particularly in continuous-valued state spaces. 

To address the limitations of Q-learning in high-dimensional and continuous state spaces, Deep Reinforcement Learning (DRL) has emerged as a powerful technique. DRL is a type of RL that leverages deep neural networks to approximate the state-action-reward relationships instead of explicitly storing Q-values for individual states. The seminal paper \cite{mnih2015human} introduced the first implementation of DRL, employing two Deep Neural Networks (DNNs). One network captures the complex state-action-reward relations by mapping states to actions, while the other network generates optimal actions based on the learned relationships. The learned modeling network is flushed to the action-generating network, once in a while.

\section{Problem Formulation}
\label{sec:problem}

A core part of our optimization is heat transfer modeling in open-office and mixed settings. 
Generally, the heat transfer characteristics are influenced by factors like types of he materials used in construction, the presence of insulation, and temperature gradients between spaces \hw{ \cite{xiong2021study}}.
Nevertheless, the thermal behavior of closed and open office spaces can be substantially different and influenced by the unique design features of each type. For instance, in closed offices, the presence of solid walls blocks heat transfer through airflow circulations, so heat transfer is primarily performed through conduction, which follows a different set of physics rules, as opposed to open offices where heat transfer is mainly through convection. In mixed settings, when some partitioning walls are not of full height, both heat transfer modes coexist making the direct analysis even more complex due to a more intricate temperature distribution within the space. It also challenges individual zone temperature control since the air flow diffuses the heat to neighbor zones. To accommodate such conditions, we use a multi-input, multi-output DRL model for its flexibility in modeling complex relations.

\subsection{Multi-Zone Thermodynamics }

\begin{table}[h]
\centering
\caption{Notation used for thermal model}
\label{tab:notation}
\resizebox{1\columnwidth}{!}{%
\begin{tabular}{|l|l|l|} 
\toprule
\multicolumn{1}{|c|}{Symbol} & \multicolumn{1}{c|}{Definition}                            & \multicolumn{1}{c|}{Units}  \\ 
\hline
$A_{win},i$                    & Area of the window in room/zone i                          & $m^2$                       \\ 
\hline
$\alpha_{win},i$               & Solar absorptance of the windows in room/zone i            & \%                          \\ 
\hline
$T_{sol},i$                    & Solar temperature on the windows in room/zone i            & K                           \\ 
\hline
$T_i$                        & Air temperature in room/zone i                             & K                           \\ 
\hline
$N_i$                        & Set of adjacent rooms/zones to room/zone i                 & Count                       \\ 
\hline
$k_i,j$                      & Thermal conductance between rooms i and j                  & W/(m$^2$K)                  \\ 
\hline
$A_i,j$                      & Wall/surface area between rooms i and j                    & $m^2$                       \\ 
\hline
$d_i,j$                      & Wall thickness between rooms i and j                       & $m$                         \\ 
\hline
$h_i,j$                      & Convective heat transfer coefficient between zones i and j & W/(m$^2$K)                  \\ 
\hline
$\dot{m_i}$                  & Flow rate of the air mass supplied by HVAC i               & kg/s                        \\ 
\hline
$C_p,i$                      & Specific heat capacity of air in room i                    & J/(kg$\cdot$K)              \\ 
\hline
$T_{hvac},i$                   & Supply air temperature of the HVAC in room/zone i          & K                           \\
\bottomrule
\end{tabular}}
\end{table}

Suppose there exist $n$ rooms in a conventional closed office, represented by $X = \{ x_1,x_2,...x_N\}$.
The heat gain of room $x_i$ in such an office can be expressed as \cite{nagarathinam2015centralized, nagarathinam2017energy}:

\begin{align}
\label{eq:t1}
\nonumber
\Delta Q_{x_i} = & Q_{int,x_i} + Q_{solar,x_i} + Q_{cond,x_i} + Q_{hvac,x_i} \\
 \nonumber
 = & Q_{int,x_i} \\
& +\alpha_{win,x_i} A_{win,x_i} \left(T_{sol,x_i}^4 - T_{x_i}^4\right) \\
 \nonumber
& + \sum_{j \in N_i} ( \frac {k_{x_i,x_j}}{d_{x_i,x_j}} A_{x_i,x_j}  (T_{x_j} - T_{x_i})) \\
& + \dot{m_{x_i}} C_{p,x_i} (T_{hvac,x_i} - T_{x_i}),
\end{align} 
where $\Delta Q_{x_i}$ represents the thermal energy change in room $x_i$,
comprising four 
components, (i) the internal heat gain due to lighting, equipment, and occupants $Q_{int,x_i}$, (ii) the solar energy gain $Q_{solar,i}$, (iii) the energy generated by HVAC system $Q_{hvac,i}$, and (iv) the heat loss/gain through conduction $Q_{cond,i}$. In this equation, $N_i$ is the set of adjacent zones to $x_i$. A complete list of parameters and their units can be found in Table \ref{tab:notation}.

On the other hand, suppose there exist $n$ connected spaces $Y = \{ y_1,y_2,...y_n \}$ in an open-office design, where there is no conduction-based heat transfer through interior walls. Instead, heat is mainly distributed to neighboring spaces through convection:

\begin{equation}
\begin{aligned}\label{eq:t2}
Q'_{conv,y_i} = \sum_{j \in N_i} h_{y_i,y_j} A_{y_i,y_j} (T_{y_j} - T_{y_i}),
\end{aligned} 
\end{equation}
where $h_{i,j}$ is the convective heat transfer coefficient between zones i and j, $A_{i,j}$ is the surface area between zones i and j. Note that in (\ref{eq:t1}) and (\ref{eq:t2}), $T_i$ is the temperature of zone $x_i$, by modeling each zone as a point object. This approximation is more appropriate for the steady-state situation and can serve only as an approximation for the transient time when different parts of the zone may have different temperatures. However, this is not a concern since our DRL model is flexible enough to model and compensate for such second-order terms. 
By replacing (\ref{eq:t2}) in (\ref{eq:t1}), we obtain

\begin{equation}
\begin{aligned}\label{eq:ty}
\Delta Q_{y_i} =& Q_{int,y_i} + Q_{solar,y_i} +Q'_{conv,y_i} + Q_{hvac,y_i}  \\
=& Q_{int,y_i} \\
& +\alpha_{win,y_i} A_{win,y_i} \left(T_{sol,y_i}^4 - T_{y_i}^4\right) \\
& + \sum_{j \in N_i} h_{y_i,y_j} A_{y_i,y_j} (T_{y_j} - T_{y_i})\\
& + \dot{m_{y_i}} C_{p,y_i} (T_{hvac,y_i} - T_{y_i}).
\end{aligned} 
\end{equation}

To further simplify the heat transfer analysis and isolate the effects of building design, we assume that offices $x_i$ and $y_i$ 
are located in an ideal environment, where there is no solar radiation or occupant-generated heat, and the initial energy of systems is zero. Therefore, (\ref{eq:t1}) and (\ref{eq:ty}) reduce to:

\begin{align}
\label{eq:x-approx} 
\nonumber
\Delta \hat{Q}_{x_i} = & \sum_{j \in N_i} ( \frac {k_{x_i,x_j}}{d_{x_i,x_j}} A_{x_i,x_j}  (T_{x_j} - T_{x_i})) \\
& + \dot{m_{x_i}} C_{p,x_i} (T_{hvac,x_i} - T_{x_i})\\
\nonumber 
\label{eq:y-approx} 
\Delta \hat{Q}_{y_i} =& \sum_{j \in N_i} h_{y_i,y_j} A_{y_i,y_j} (T_{y_j} - T_{y_i})\\
& + \dot{m_{y_i}} C_{p,y_i} (T_{hvac,y_i} - T_{y_i})
\end{align}

According to the first law of thermodynamics, we have:

\begin{equation}
\begin{aligned}\label{eq:td1}
\Delta{Q} = {Q_{in}} - {Q_{out}},
\end{aligned} 
\end{equation}
where $\Delta Q$ is the office heat change,  ${Q_{in}}$ is the heat gained by the office, and ${Q_{out}}$ is the heat lost by the office.
In this case, heat gain in offices can only come from HVAC equipment. Therefore, in steady-state conditions, we have:

\begin{align}
\label{eq:x-last}
\dot{m_{x_i}} C_{p,x_i} (T_{hvac,x_i} - T_{x_i}) 
&= - \sum_{j \in N_i} ( \frac {k_{x_i,x_j}}{d_{x_i,x_j}} A_{x_i,x_j}  (T_{x_j} - T_{x_i}))\\
\label{eq:y-last}
\dot{m_{y_i}} C_{p,y_i} (T_{hvac,y_i} - T_{y_i}) 
&= - \sum_{j \in N_i} h_{y_i,y_j} A_{y_i,y_j} (T_{y_j} - T_{y_i}) 
\end{align}

It can be seen that the thermodynamics of closed offices and open offices exhibit substantial differences, meaning that the prior models that consider closed offices can not be directly transferred to open-office setups. 
In other words, for closed offices, the heat gain for each room primarily depends on its dedicated HVAC equipment and the heat transfer occurring through conduction between spaces. On the other hand, open offices operate on a different principle. The heat gain in an open office zone is influenced by the HVAC units in its own area as well as those in neighboring areas. Heat transfer in open offices predominantly occurs through convection, with air circulation playing a crucial role. Our multi-input multi-output DRL model has the flexibility of modeling open, closed, and mixed setups.

\begin{figure}[h]
    \centering
    \centerline{\includegraphics[width=0.6\columnwidth]{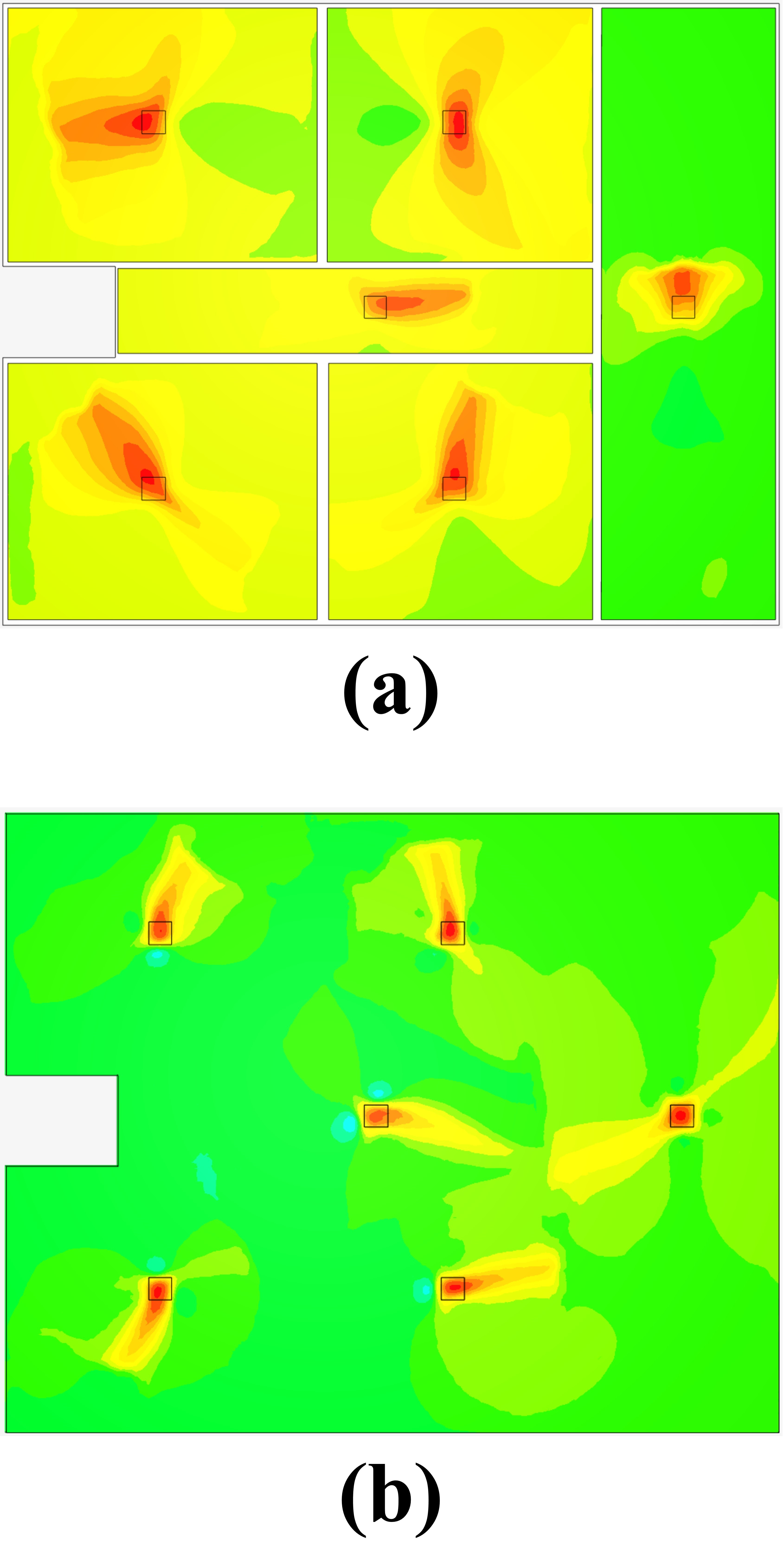}}
    \caption{
    Heat transfer comparison in (a) offices with closed configuration versus (b) offices with open-plan configuration.}
    \label{fig:Open_Office}
\end{figure}

To validate our assumptions regarding the differences in heat transfer between closed and open office spaces, we employed SimScale to simulate the conjugate heat transfer of the same building once with a closed plan and next, with an open plan. SimScale is a cloud-based simulation platform that can model airflow using Computational Fluid Dynamics (CFD) and Finite Element Analysis (FEA) analysis.  
The results in Figure \ref{fig:Open_Office} show a more even and smoother temperature distribution for an open office configuration compared to the sharp transitions in borders between different zones in a closed setting, as expected.
These distinctions highlight the need for energy management strategies that are applicable to open-plan offices.

\subsection{Thermal Comfort Metric}

\ifx \myver \hugever

\sout{There exist multiple metrics to evaluate occupant comfort in buildings. Among them is the Predicted Percentage of Dissatisfied (PPD) metric, which predicts the percentage of occupants who are likely to be dissatisfied with the thermal environment based on the ASHRAE Standard 55 guidelines. PPD can be used by building energy managers to improve thermal comfort and reduce discomfort and related health and productivity issues; However, its calculation process is over-complicated \ar{for involving occupant interview/opinion?!!!} and incorporating many factors such as air temperature, mean radiant temperature, relative humidity, air velocity, clothing, the activity level of occupants, and other implicit factors \cite{valladares2019energy}.
Some prior works including \cite{deng2022towards} utilized a simplified PPD calculation model, but it may not be suitable for this study due to differences in the environment, climate zone, and building functionality.}

Some comfort metrics like Predicted Percentage of Dissatisfied (PPD) incorporate many factors such as air temperature, mean radiant temperature, relative humidity, air velocity, clothing, the activity level of occupants, and other implicit factors to calculate occupant comfort \cite{valladares2019energy}. 

\fi

In our study, 
thermal comfort is simply defined as keeping the current temperature of zone $i$ within a certain range that makes the occupants feel comfortable. In other words, we define the temperature violation or \textbf{Comfort Compliance Ratio (CCR)} 
as
\begin{align}
    CCR &= \frac{1}{ N T} \sum_{i=1}^{N}\int_{t=0}^{T} I(T^{min}_i \leq T_{i,t} \leq T^{max}_i) ~dt,
\end{align}
where $T$ is the test period, $T_{i,t}$ is the temperature of zone $i$ in time $t$, $I()$ is the indicator function, $N$ is the number of zones, and $[T^{min}_i,T^{max}_i]$ is the comfortable temperature range for zone $i$ \cite{ASHRAE201755}. This metric can be seen as the compliance ratio of occupants averaged over all zones $x_1$ to $x_N$. For a discrete-time system with $t=1,2,3,\cdots,T$, it can be simplified to:
\begin{align}
    CCR &= \frac{1}{ N T} \sum_{i=1}^{N} 
 \sum_{t=1}^{T} I(T^{min}_i \leq T_{i,t} \leq T^{max}_i). 
\end{align}

Equivalently, we define the Comfort-Violation Ratio (CVR) as $CVR=1-CCR$. 
The goal is to maintain maximal comfort-compliance.

To assess the effectiveness of the proposed method in this study, we compared our method against the currently active 
RBC method, from two key perspectives.
Firstly, we utilized the comfort violation metric to assess the level of thermal comfort achieved by the proposed DRL model. This metric allowed us to quantify how the proposed DRL approach can maintain desired temperature ranges and minimize temperature deviations. 
Secondly, we evaluate the trade-off between energy saving and comfort loss by adjusting the coefficient of the comfort violation, to explore how the proposed comfort metric influence the energy optimization task.
The results of this experiment are provided in Section \ref{sec:trade}.

\begin{figure}[htbp]
    \centering
    \centerline{\includegraphics[width=0.6\columnwidth]{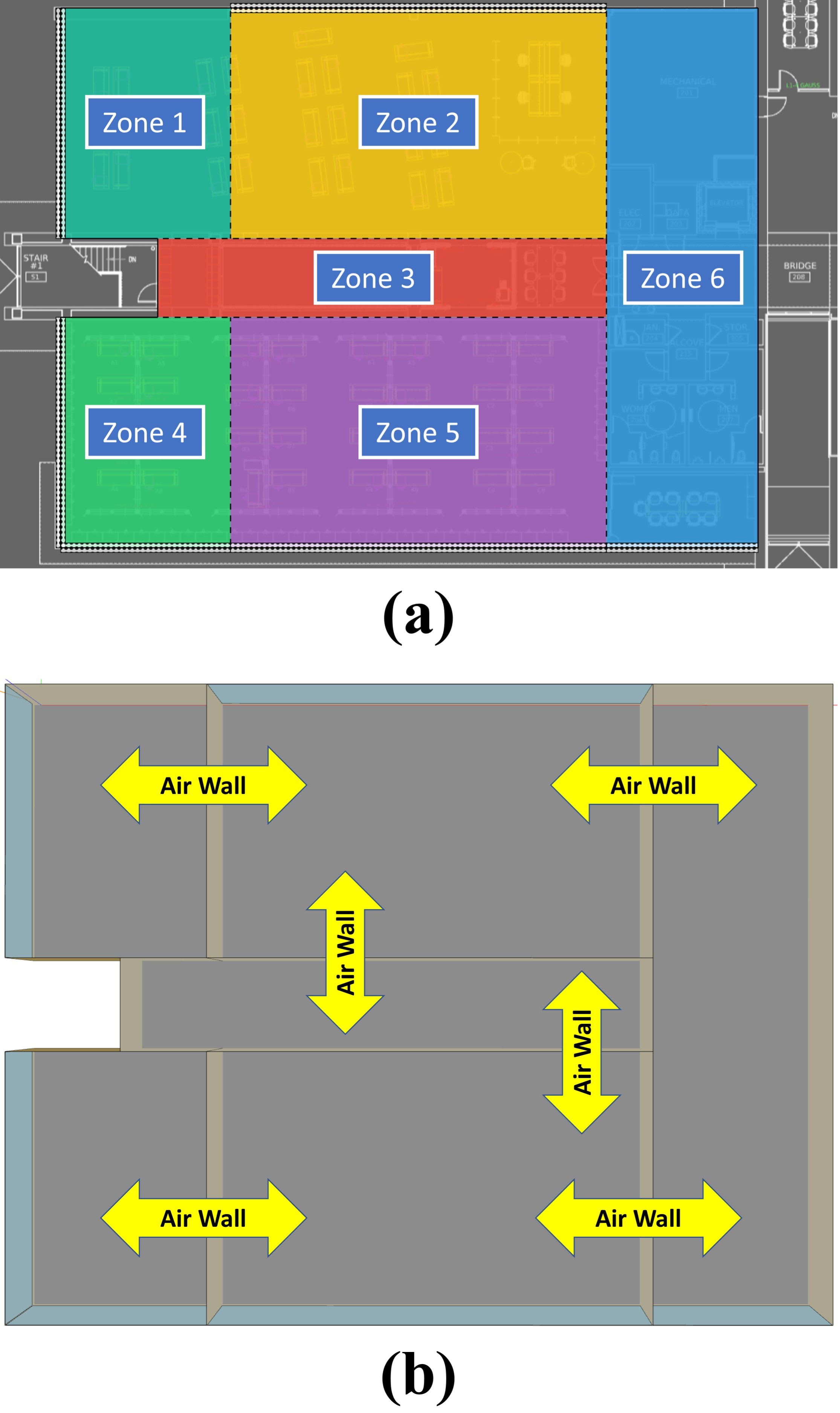}}
    \caption{
    Floor plan design. (a) An open office with six-thermal zones; (b) Air Wall modification using Python.}
    \label{fig:floorplan}
\end{figure}

\section{Proposed Solution}
\label{sec:solution}

\subsection{Open Office Model Design}
\label{sec:office_model}

In this paper, we focus on the open floor configuration, where a single space is divided into multiple thermal zones that are connected without physical wall isolation. 
We use OpenStudio to develop an open office model based on a real building. To precisely represent the interconnected nature of the thermal zones within the open office, we developed a custom Python function that can modify the surface material of certain areas to 'air', as shown in Figure \ref{fig:floorplan}. This logical partitioning allows us to study the energy optimization and thermal management of interconnected zones within a single space.

\begin{figure}[htbp]
    \centering
    \centerline{\includegraphics[width=1\columnwidth]{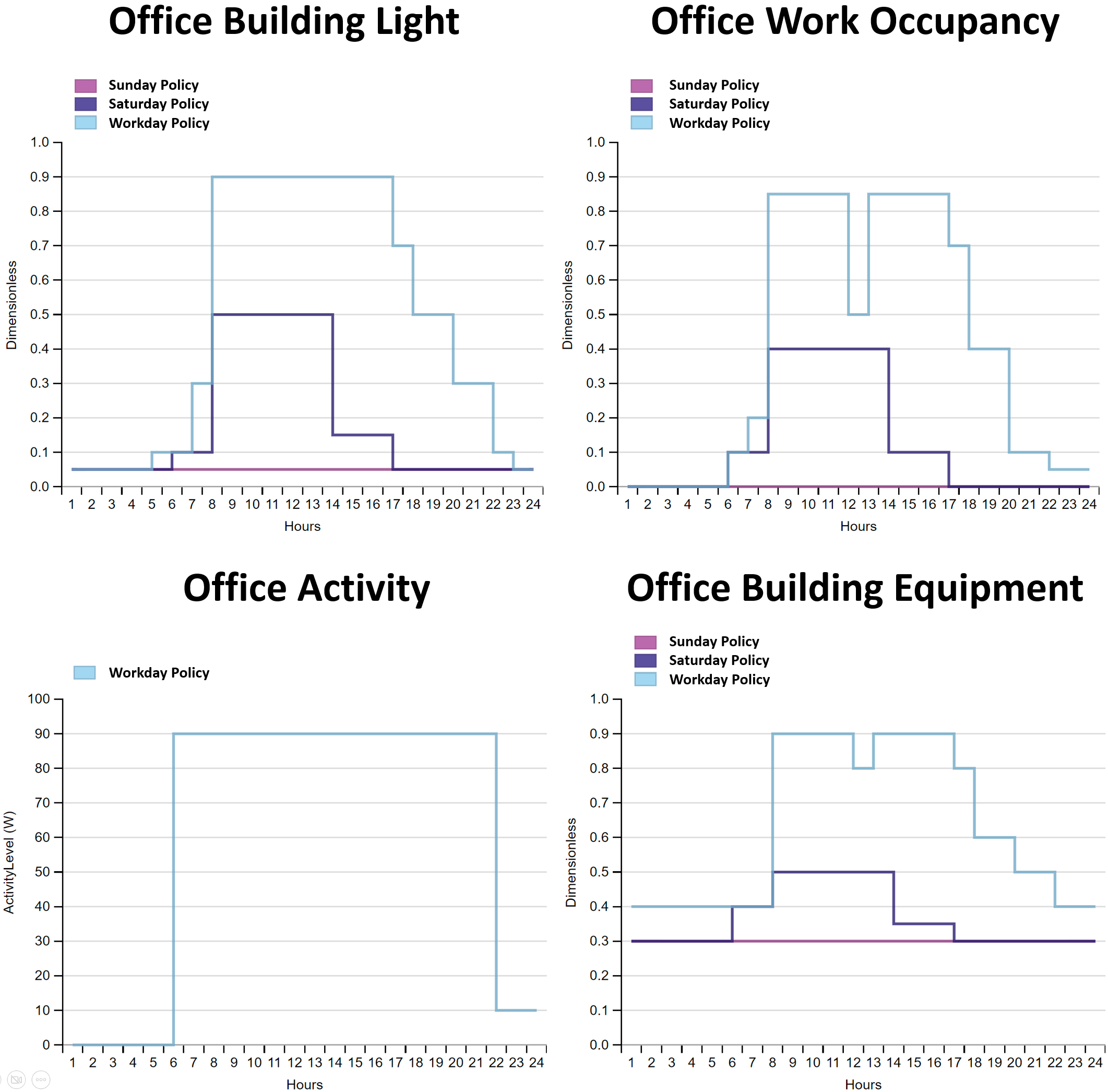}}
    \caption{
    \ar{Simulated Office Schedule and Activity Level. }}
    \label{fig:occ}
\end{figure}

\hw{Additionally, we applied various load factors, including the office occupancy rate, the work schedule, electronic equipment load, and the light schedule to maximally mimic the selected building's real-world conditions, as shown in Figure \ref{fig:occ}.}

\subsection{System States}

The primary control variables considered in our approach (i.e. the actions in the DRL model) are the setpoints of the thermostats associated with each individual zone, which directly influenced the operation of their respective VAV unit.

\begin{table}[h]
\centering
\caption{Variables for DQN Input (States)}
\label{tab:key_value}
\resizebox{1\columnwidth}{!}{%
\begin{tabular}{lll}
\toprule
\textbf{Variable} & \textbf{Definition}             & \textbf{Values (Unit)} \\
\midrule
\textbf{O}      & Outdoor Dry AirBulb Temperature & 25 $\sim 110 ({}^\circ{F})$              \\
\textbf{T}        & Zone Temperature                & 60 $\sim 90 ({}^\circ{F})$              \\
 \textbf{V}        & Zone VAV Status                      &0, 1              \\
\textbf{W}       & Active Working Time             & 0, 1                 
\\
\bottomrule
& & \end{tabular}}
\end{table}

The system state is defined as $S_t = \{ O_t, T_{i,t}, W_t, V_{i,t} \}$ for $i-1,2,\cdots,6$, where
$O_t$ is the outdoor air dry bulb temperature, 
$T_{i,t}$ is the indoor temperature of zone $x_i$, 
$W_t$ is the work time indicator, 
and 
\hw{$V_{i,t}$ is the status of the VAV unit in zone $x_i$, all evaluated at time $t$.}
Therefore, the state vector (the input to the DQN network) has a total of 14 variables, $2$ global environment variables ($O_t$,$W_t$), and $12$ zone-specific variables ($T_{i,t}, V_{i,t}$, 2 per zone). The range and unit of these parameters are given in Table. \ref{tab:key_value}.

\subsection{HVAC Control Actions}

\begin{figure}[h]
    \centering
    \centerline{\includegraphics[width=1\columnwidth]{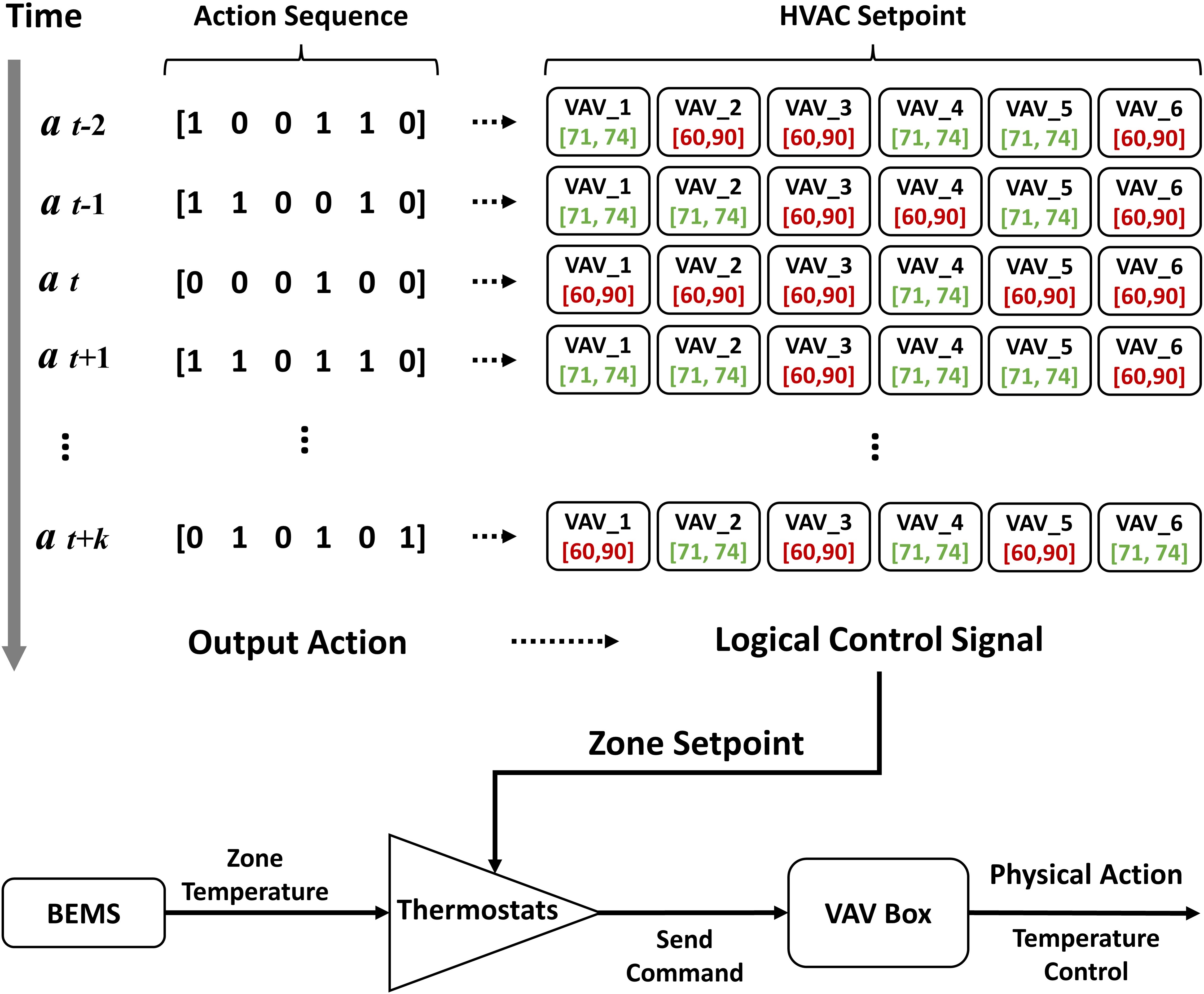}}
    \caption{Control action translates to physical command for HVAC equipment.}
    \label{fig:action}
\end{figure}

Each thermal zone contains a VAV unit with heating and cooling functions. Compared to the conventional DRL approaches (e.g., the method proposed in \cite{wei2017deep}) that define the actions of each control unit by the actual temperature range of the thermostat, we take actions in two sequential steps. First, we take a \textit{logical action} that includes two possibilities for each VAV: \textbf{Comfort Policy ON} ($a=1$), and \textbf{Comfort Policy OFF} ($a=0$). We present these two states with \textbf{CCR ON} and \textbf{CCR OFF} in the rest of this paper. Turning on and off the Comfort Policy determines the set temperature range as $[71\sim74]{}^\circ{F}$ and $[60\sim90] {}^\circ{F}$ \cite{ASHRAE201755, ASHRAE202262}. Then, the \textit{logical action} along with the current \textit{zone temperature} determines the \textit{physical action} to be taken by turning on and off the heater and cooler systems. For example, if $T_{i,t}=68$ in the \textit{CCR ON} mode, then the system turns on the heater for zone $x_i$, whereas it takes no action under \textit{CCR OFF} mode. The relationship between the \textit{logical} and \textit{physical} actions is demonstrated in Figure \ref{fig:action}. 
Note that the output of the DQN network determines the \textit{logical action} which translates to setting the allowable temperature range, then the \textit{physical actions} are taken by the BEMS. Therefore, the action space for 6 thermal zones is $2^6 = 64$.

The Q network only predicts the Q value of the \textit{logical action} $a \in \mathcal{A} = \{0,1\}$. This approach helps the Q network converge much faster than the conventional way of setting an exact temperature setpoint for the next time step.

\begin{figure*}[h]
    \centering
    \centerline{\includegraphics[width=0.8\textwidth]{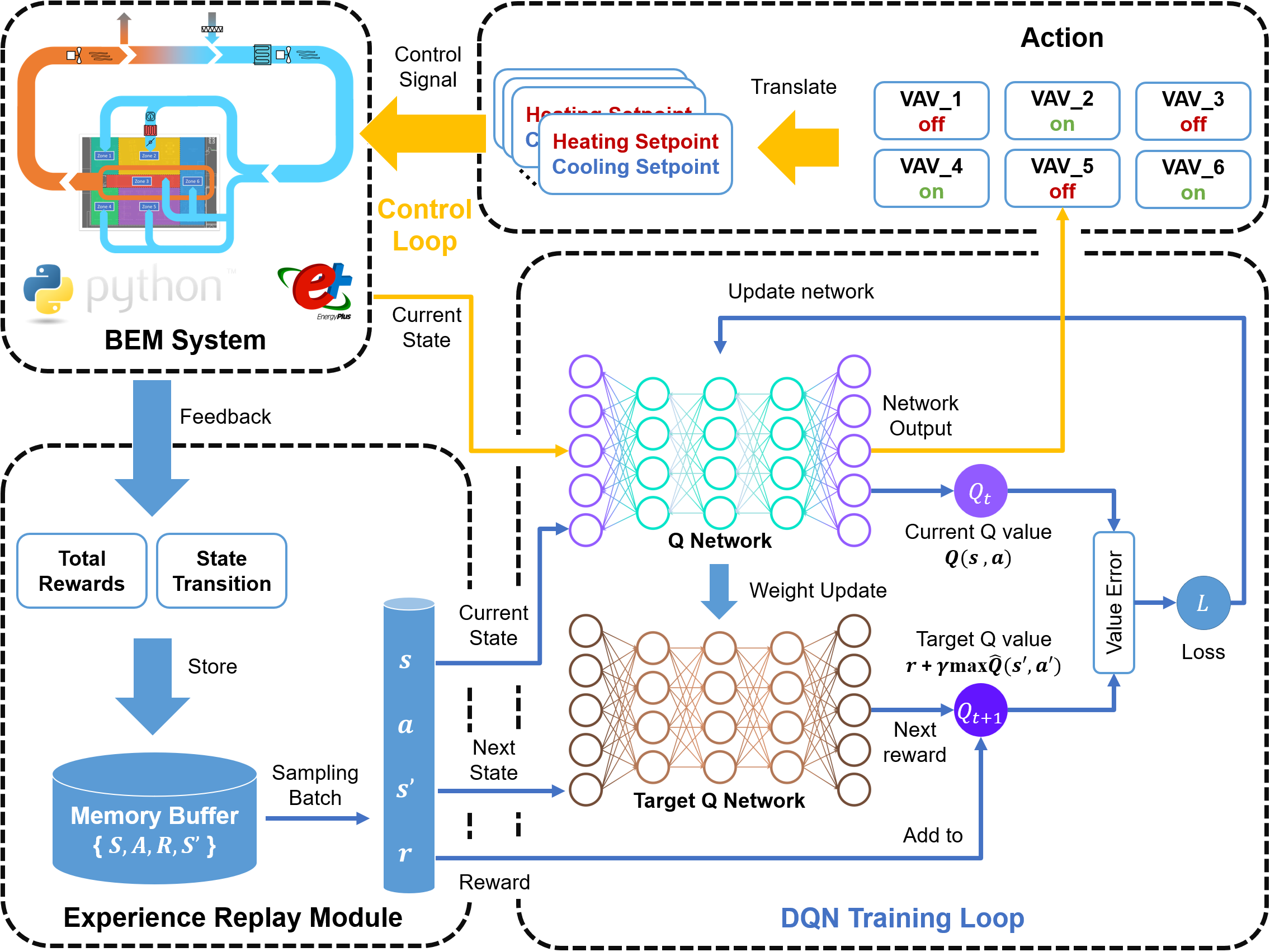}}
    \caption{\hw{Training flow of implemented DQN-based framework.}}
    \label{fig:training_flow}
\end{figure*}

\subsubsection{Reward Mechanism through Loss Function} \label{sec:reward}
\label{sec:reward}

In this paper, we define the reward mechanism through a loss function of the DQN network to achieve both energy optimization and thermal comfort. In the meantime, the proposed reward function also considers a few constraints accommodating real-world requirements. For instance, setting reasonable upper and lower bounds for HVAC setpoints to prevent health or safety issues (in the \textit{CCR OFF} mode), and incorporating a term to penalize frequent ON/OFF transitions to reduce the mechanical wear of the ventilation system.

More specifically, the loss function has three components \hw{ $\mathcal{R}=- \mathcal{L}_{total}=
-( \mathcal{L}_{C}+\mathcal{L}_{E}+\mathcal{L}_{S}$}) with the following details. 
\hw{The first term is defined as 
\begin{align}\label{eq:LT}
\mathcal{L}_C = \sum_{i}^{n} \mathcal{L}_{T_i} + \sum_{i}^{n} \mathcal{L}_{H_i},
\end{align}
which is the summation of zone-specific terms to enforce thermal comfort. Specifically, $\mathcal{L}_{T}$ represents the temperature loss and $\mathcal{L}_{H}$ represents the humidity loss. Specifically, }
\begin{align}\label{eq:LTi}
\mathcal{L}_{T_i} = \begin{cases}
0 & \text{if } T_{\text{min}} \leq T_i \leq T_{\text{max}}, \\
\eta_T * ({T_i}-{T_{min}})^2, & \text{if } T_i 
 \leq T_{\text{min}} \\
\eta_T * ({T_i}-{T_{max}})^2, & \text{if } T_i \geq T_{\text{max}} \\
\end{cases},
\end{align}
 $T_i$ is the temperature of zone $x_i$, $T_{target}=(T_{\text{min}} +T_{\text{max}})/2$ is the mean of the thermal comfort range, and $\eta_T$ is a tunable temperature loss factor. According to (\ref{eq:LT}) and (\ref{eq:LTi}), a positive reward is collected if the current temperature is within the desired range, and a negative reward, proportional to the distance to the center of the desirable range is considered when the temperature is off the range. 

This formulation is inspired by the fact that the optimum temperature to maintain maximal working efficiency and personnel comfort is not a solid number but an implicit range \cite{ASHRAE202262}. Therefore, we set the loss proportional to the distance between the current temperature and the mean of the target temperature range (${T_i}-{T_{target}}$). 

Note that our heuristic loss formulation is different than the more commonly used comfort loss function in \hw{other} DRL approaches defined as
\begin{align}\label{eq:LTi}
\mathcal{L}_{T_i}' = \begin{cases}
0, & \text{if } T_{\text{min}} \leq T_i \leq T_{\text{max}}, \\
1, & \text{if } T_i 
 \leq T_{\text{min}} \text{or } T_i \geq T_{\text{max}}, \\
\end{cases},
\end{align}
which equally penalizes off-range temperatures. This helps the DQN converge faster since the proposed loss function can inform the precise distance from the goal.

This loss formulation is considered only at working hours or in \textit{CCR ON} mode. At all times, including off-work hours, we restrict the temperature to never go beyond a reasonable safe range ($60{}^\circ{F} - 90{}^\circ{F}$) to maintain safety for equipment (servers, computers, appliances, etc.) according to ANSI/ASHRAE Standard 62.1-2022 \cite{ASHRAE202262}.
We set this as a hard constraint in our code, but it can also be implicitly imposed by assigning an infinitive negative reward when the temperature is off the safe zone by adding the following line to (\ref{eq:LTi}). 
\begin{align}\label{eq:LTi2}
\mathcal{L}_{T_i} = \infty & \text{   if } 60^\circ < T_i \text{ or } T_i > 90^\circ 
\end{align}

\hw{
Meanwhile, the humidity loss $\mathcal{L}_{H_i}$ is defined as
\begin{align}\label{eq:LTi}
\mathcal{L}_{H_i} = \begin{cases}
0 & \text{if } H_{\text{min}} \leq H_i \leq H_{\text{max}}, \\
\eta_H * |{H_i}-{H_{min}}| & \text{if } H_i \leq H_{\text{min}},  \\
\eta_H * |{H_i}-{H_{max}}| & \text{if } H_i  \geq H_{\text{max}}, \\
\end{cases}
\end{align}
where ${H_i}$ represents the humidity of  zone $x_i$, and ${H_{target}}$ represents the desired humidity range.
While the desired humidity range is not a hard constraint \cite{ASHRAE202262, ASHRAE201755}, we apply a range of $20\% \sim 80\%$ that is used by many studies \cite{alahmer2013vehicular, roshan2017defining, albatayneh2021significance}.
}

Another objective of the optimization is to minimize energy consumption. RL-based methods, as opposed to rigid rule-based methods, provide more flexibility in balancing the trade-off between comfort level and energy use. To this end, the loss function includes the following term 
\begin{align}\label{eq:energy}
\mathcal{L}_E =  \eta_E * {E_{t}} 
\end{align}
where $E_{t}$ is the energy consumption of time step $t$, and $\eta_E$ is the energy loss factor. Note that we applied a post-normalization to energy loss and comfort loss since their primary scales are substantially different. 
After each step of the simulation, the system will notify the amount of HVAC electricity consumption. Note that the energy consumption of the HVAC system depends on many factors, such as the equipment brand, performance, and energy efficiency. However, it does not matter since we use post-normalization.

The last term of the loss function is 
    \begin{align}\label{eq:onoff}
    \mathcal{L}_{S} = \eta_S * \sum_{i}^{n} (A_{i,t} \oplus A_{i,t-1})
    \end{align} 
\hw{to impose \textit{smoothness}, where $\oplus$ is XOR operation, $A_{i,t}$} is the action of VAV uni $i$ at time step $t$, and $\eta_S$ is the \textit{smoothness} loss factor. This term is used to penalize unnecessary on/off transitions that cause unnecessary discomfort \ar{and undermine the system's energy efficiency, noting that the system uses more energy during transition intervals.}

We use \textit{Boolean XOR} to calculate transition loss between our binary \textit{logical actions} per VAV, as shown in Figure \ref{fig:action}.

\subsection{DRL Model Design}

The role of deep learning in DRL is two-fold. The primary deep Q network is used to estimate the Q-values for state-action combinations while the target network is used to predict Q values for the next action. 
To construct our DRL model, we employ a modified twin-deep Q-network implementation \cite{wei2017deep}. Compared to other advanced DRL models such as Deep Deterministic Policy Gradient (DDPG) \cite{du2021intelligent} and Multi-Agent Deep Reinforcement Learning (MADRL) \cite{yu2020multi, fu2022optimal}, this proposed DQN benefits from generalizability and shorter training time.

\begin{table}[h]
\centering
\caption{\hw{Hyperparameter Setting For Training}}
\label{tab:paras}
\resizebox{1\columnwidth}{!}{%
\begin{tabular}{llll} 
\toprule
\textbf{Hyperparameter} & \textbf{Definition}                               & \textbf{Value}  & \textbf{Unit}\\
\midrule
Epochs& Default training epochs& 20              &\hw{epoch}\\
lr                      & Learning rate& 0.001           &-\\
Gamma& Discount factor& 0.9             &-\\
Epsilon& Greedy factor \ar{($\epsilon$)}& 0.1               &-\\
Buffer Size& Size of the memory buffer& 10,000          &-\\
Minimal Size& Minimal time interval to sample the memory buffer& 200             & \hw{time steps}\\
Batch Size& Batch size of sampled memories& 128             & \hw{samples}\\
Target Update& The time interval for target Q network updating& 20              &\hw{time steps}\\
State Dim& Input dimension of Q network& 14              &-\\
Action Dim& Output dimension of Q network& 64              &-\\
Network Layer Num& Number of hidden layers of the Q network& 3               &-\\
Hidden layer neuron& Number of neurons for each hidden layer& 128&-  

\\
 Energy factor& The factor of energy penalty $\eta_E$& 5&-\\
 Temperature factor& The factor of temperature violation penalty $\eta_T$& 5&-\\
 \ar{Humidity factor}& \ar{The factor of humidity violation penalty} \ar{$\eta_H$}& \ar{1}&\ar{-}\\
 Signal factor& The factor of signal smoothness penalty $\eta_S$& 1&-\\
 \bottomrule
 & & &\\
 \end{tabular}}  

\end{table}

As shown in Figure \ref{fig:training_flow}, the training process of the developed open office model involves the interaction between the EnergyPlus simulation program and the action producer program in Python which acts as the agent of the DRL. The transition set $\{ s_t, a_t, r_t, s_{t+1}\}$ is stored in the memory buffer (to be used for experience replay). A mini-batch is then sampled and fed into the primary Q network, while the target Q network takes the next state as input and estimates the Q value of the next action $Q(s_{t+1},a_{t+1})$ (also shown as $Q(s',a')$ in some literature). The loss is calculated by comparing this estimated Q value with the Q value of the current state  \hw{$Q_t(s,a) = r_t(s,a)+\gamma \max\limits_{a'} Q_t(s',a')$},
where $\gamma$ is the discount factor \hw{and $r_t(s,a)$ is the reward $\mathcal{R}$}. The gradient of the loss \hw{$\mathcal{L} = (Q_{target}-Q_{current} )^2 = \|[r_t(s,a)+\gamma \max\limits_{a'} Q_t(s',a')] - Q_t(s,a)  \|^2$} is used to update the \hw{current Q network}. The weights of the Q network are flushed to the target Q network, once every 200 iterations. 
The system proceeds by inputting the current state into the target Q network and selecting the action with the maximum Q value. The actor module takes this action as input, transitions to the next state, and sends the resulting transition back to the Python-based DRL agent. This training loop continues until the end of the simulation, ensuring the gradual improvement of the model. 
After training is completed, the trained DQN is used to produce the best actions based on the learned optimal policy under any condition.
Table \ref{tab:paras} presents the list of hyperparameters used for the DQN training.

\section{Case Study}
\label{sec:case}

The building model in this work is a simplified model of a real-world case, located in Greenville, SC. The building contains $3$ stories and each floor consists of two-large open offices and other closed rooms such as conference rooms and closed offices. 
The size of the target open office is $10,027 ft^2$ in total and has been divided into six different thermal zones with air walls. Each zone has its own thermal control unit since the real building equipped a VAV box for their designed thermal zones.




\begin{table}[h]
\centering
\caption{\hw{Environment Setting for Simulation.}}
\label{tab:env_set}
\resizebox{1\columnwidth}{!}{%
\begin{tabular}{llll}
\toprule
\textbf{Variable}            & \textbf{Definition}                            & \textbf{Value}              & \textbf{Unit}        \\
\midrule
Air Wall            & Outdoor dry bulb air temperature      & [True, False]& -           \\
Roof Sun            & Is the building roof exposed to the sun& [True, False]& -           \\
Air Infiltration    & Is building naturally ventilated& [True, False]& -           \\
Thermal Zone        & Thermal zone division                 & [0,1,2,3,4,5,6]& -           \\
Outdoor Temperature & Outdoor temperature& 32 $\sim$ 100& $^\circ{F}$ \\
 \hw{Outdoor Humidity}& \hw{Outdoor humidity}& \hw{0$\sim$100}&\hw{\%}\\
 Time Range& Start time and end time of simulation & 01/01 $\sim$ 12/31&Day         \\
 Time Step& Simulation steps per hour             & [1,3,6,12]&-           \\
Time Interval       & Simulation interval per hour          & [60,20,12,5]& min         
\\
\bottomrule

& & & \\\end{tabular}}
\end{table}

We used Openstudio \cite{NREL2022openstudio} to build an open office model that is located on the 2nd floor of the case building, as discussed in Section \ref{sec:office_model}. Then, we used EnergyPlus Python API (22.0) to simulate the energy consumption of the proposed open office. EnergyPlus is a simulating core that is defined by the Department of Energy (DOE) \cite{doe2022energyplus} and has been widely used in building simulation and analysis for its advantages such as accurate simulation, computational efficiency, and model scalability \cite{doe2022energyplusengineering}.

It is imperative to note that the \textit{air infiltration} of the building model in this work has been set to \textbf{False}, as most modern commercial buildings are fully sealed to ensure efficient air conditioning and thermal performance. This is in contrast to building models in prior research works that incorporate natural ventilation through windows \cite{nagarathinam2017energy, dai2023deciphering}. Additionally, the \textit{sun exposure} to the building roof is set to \textbf{False} since the simulated open office is located on the 2nd floor.

During each simulation time interval, the EnergyPlus Python API will generate an array of variables including the current outdoor temperature, the current temperature of each zone, the work hour indicator, and the current setpoint of each VAV unit. The proposed deep Q network will take these variables as the input vector and then output the control actions. The control actions will be translated into physical control signals for each VAV unit, and sent back to the EnergyPlus function to execute.

\ifx \myver \hugever
Among a variety of simulators, 
\arr{add other options we previously explo}, including Openstudio, and EnergyPlus,\cite{XXX}, we use EnergyPlus as our simulation platform, which is defined by the Department of Energy (DOE) 
\ar{\cite{XXX}} and has been widely used in building simulation and analysis 
\ar{for its advantages like XXX, and YYY \cite{XXX}} 
\arr{mention its key advantages and features if you know, thing like its compatibility with more enterprises, its flexibility for scripting, its more accurate results, XXXX. }.
\fi

\begin{figure*}[h]
    \centering
    \centerline{\includegraphics[width=1\textwidth]{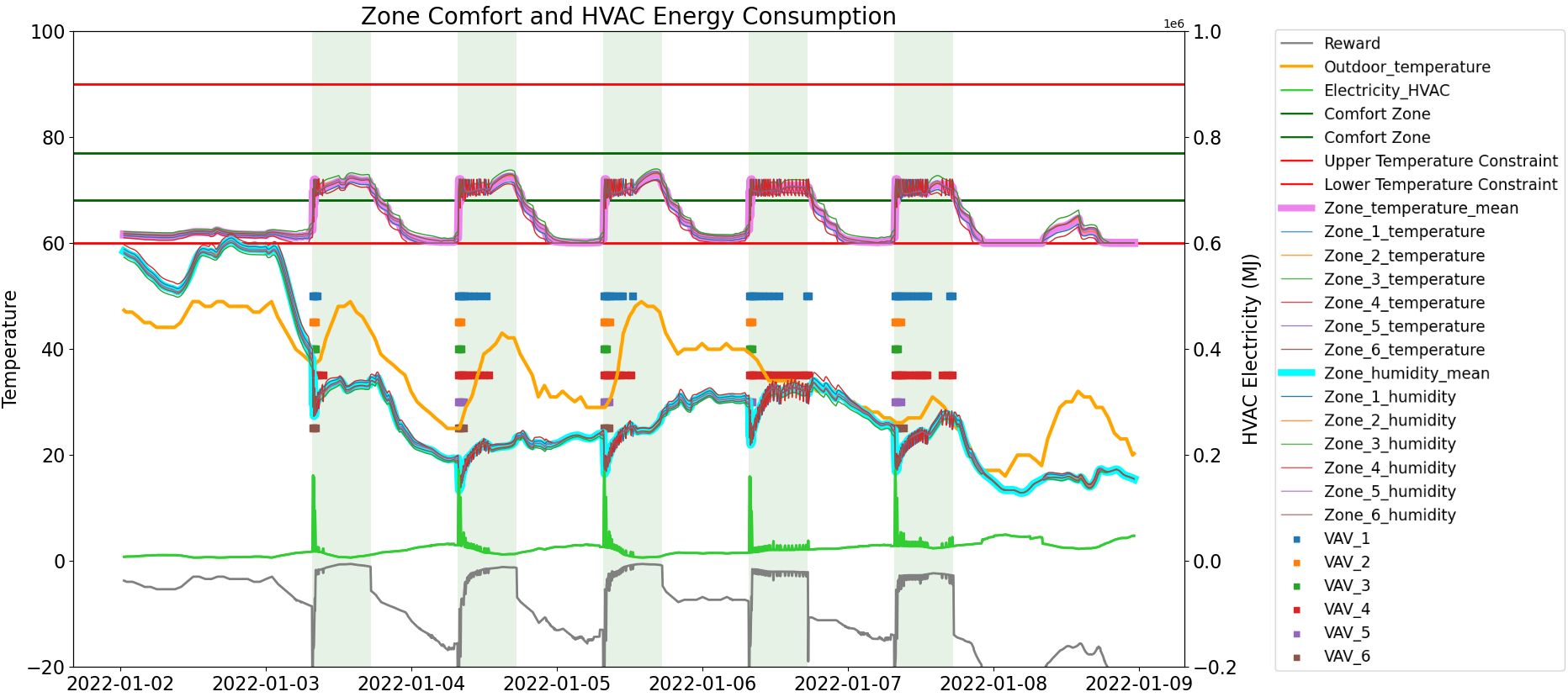}}
    \caption{\hw{Demonstration of DQN training process for one week. Top: outdoor temperature (Orange), 6 zone temperatures, temperature constraint (Red), the comfort zone (Lime Green); Middle: status of 6 VAV units, the humidity of 6 zones (Cyan); Bottom: energy  consumption of HVAC (Green),  total reward (Gray); Shallow-green background: work hours indicator.}}
    \label{fig:DQN_train}
\end{figure*}

To set up the local climate conditions at the building location, we import weather data from the EnergyPlus Weather Data Library \cite{doe2022energyplusweather}. Specifically, we select the weather data file \textbf{USA-SC-Greenville-Spartanburg.Intl.AP.723120-TMY3} for the EnergyPlus simulation. This weather file provides both historical weather data for the building location and day-stamped data required for the initialization of the simulation.

We set each time step to a number out of $[60, 30, 12, 5]$ minutes, which means respectively [1,2,5,12] time steps per hour. For example, if we set the time step to $5$ minutes, then the total number of steps becomes $365 \times 24 \times 12 = 105,120$ in total for each training epoch. 
These small time steps allow the model to learn and adapt to various weather conditions that occur throughout the year. On the other hand, by training on a diverse range of days, the model can capture different patterns and trends in energy consumption and optimize its control strategies accordingly, so it can cover a wide range of situations. 
Table \ref{tab:env_set} represents the detailed list of parameters used in our simulation environment.

All experiments are carried out using a Windows machine with an Intel i9-10900F CPU, 64GB RAM, and an NVIDIA RTX 3090 GPU. The simulation is based on Energyplus Python API (Ver. 22.0) and the DRL model is implemented in Pytorch 1.10. 
With this descent hardware configuration, the average training time is about \textbf{7.75} minutes per epoch (e.g., about 40 minutes for full training with 5 epochs), which is much less than the competitor methods. 

\subsection{DQN Training Monitoring}

To get feedback from the training process, we save and plot the output variables in real time.
Detailed information on the DQN training process is shown in Figure \ref{fig:DQN_train} for one full week by presenting various parameter variations. Specifically, the curve of outdoor temperature, the indoor temperature of each zone, and the comfort band are plotted on the top of the panel. Meanwhile, the state of the comfort policy of each VAV unit is displayed with a different color in the middle of the panel. Finally, the electricity consumption of HVAC systems along with the total reward is displayed at the bottom of the panel.

As shown in Figure \ref{fig:DQN_train}, the proposed DQN learns how to manage the control policy for each VAV unit (VAV\_1 - VAV\_6) during the training. For example, on normal weekdays \hw{($2022.1.2 - 2022.1.9$)}, DQN turns on all the VAV units at the start of the work hours and turns off a few VAV units around noon. This is because all the indoor temperatures have met the comfort level, the DQN decides which VAV unit should abandon the comfort policy to save energy, and which VAV unit should keep working to maintain the current temperature.

Note that when the comfort policy is active, the desired range is $[71\sim 74]{}^\circ{F}$ (straight-dark green lines), which is much tighter than the operation range $[60\sim90]{}^\circ{F}$, so it translates to more physical actions of the system.  
In addition to weekdays versus weekends, there is an obvious difference between the work hours (8:00 am - 5:00 pm, shallow green background) and off-work hours, when the DQN prefers to turn off more VAV units to save more energy since the comfort policy is no longer active.
Moreover, the reward curve (gray) reveals how the indoor temperature (Zone\_1\_temperature - Zone\_2\_temperature) and HVAC electricity (lime curve) will impact the loss. For instance, when the indoor temperature is out of the comfort zone or when the HVAC electricity use is high, the reward curve drops significantly.

\subsection{Zone Temperature Monitoring}

\begin{figure}[h]
    \centering
    \centerline{\includegraphics[width=0.8\columnwidth]{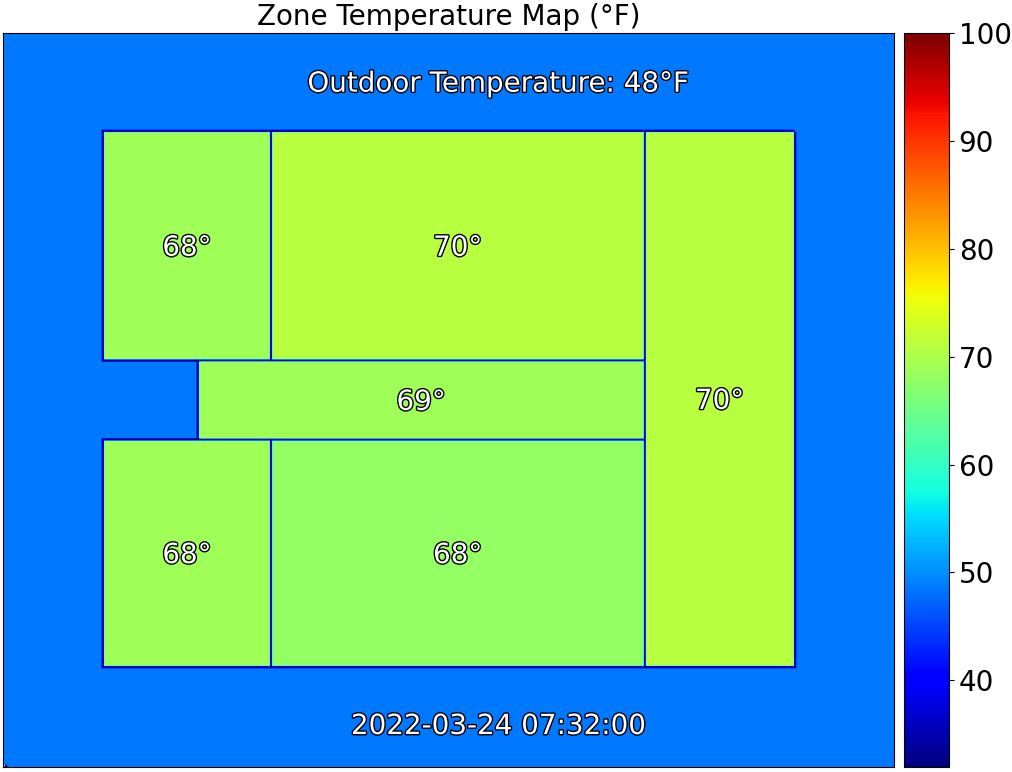}}
    \caption{Zone temperature map display during the DQN training.}
    \label{fig:zone_heatmap}
\end{figure}

To validate the feasibility of the precise temperature control for each zone in real-time, we also display a temperature map of the open office which can be displayed to the user as a computer GUI or cellphone app, as shown in Figure. \ref{fig:zone_heatmap}. 

This figure presents the average temperature of each zone in the proposed open office along with the respective color bar on the right. It is notable that the trained DQN achieves good performance in thermal comfort even when the outdoor temperature is highly inconsistent with the indoor, revealing the high potential of DQN in managing temperatures.

\section{Results}
\label{sec:results}

In this section, we demonstrate the performance of the DQN model for the proposed floor plan in terms of training efficiency, energy saving, comfort violation control, and signal smoothness control. Furthermore, we show the trade-off between the energy efficiency and comfort violation from one side and the control signal smoothness for another. 
Finally, we explore the generalizability and robustness of the proposed deep Q-network. The code of the proposed control framework is fully open-source and can be found from \ \url{https://github.com/AIS-Clemson/DRL-BEMS}.

\subsection{Standard Heat Transferring Comparison}

To establish a baseline for our comparisons, we utilize the currently used HVAC time schedule of our subject building, as the rule-based control policy in the EnergyPlus simulation. It is noteworthy that we did not apply the proposed DQN in this standard test.
Table \ref{tab:heat_trans_delta} shows the results of both closed-office and open-office configurations. We see that the HVAC system in a closed office consumes \textbf{4.89\%} more energy than the open office, consistent with our hypothesis in section \ref{sec:problem}.

\begin{table}[h]
\centering
\caption{Heat Transfer Test in Open and Close Offices.}
\label{tab:heat_trans_delta}
\resizebox{1\columnwidth}{!}{%
\begin{tabular}{cccc} 
\toprule
                                         & Closed Office & Open Office      & Unit         \\ 
\hline
Wall Materials                           & Solid         & Air              & -              \\
Energy Consumption                       & 223,151       & \textbf{212,233} & MJ           \\
Average Indoor Temperature (T)              & 69.5          & 69.45            & $^\circ{F}$  \\
Average Indoor Temperature Difference ($\Delta T$) & 1.22          & \textbf{0.47}    & $^\circ{F}$  \\
Average Indoor Temperature Variance  ($\sigma^2 T$)  & 0.69          & \textbf{0.08}    & $^\circ{F}$  \\
\bottomrule
\end{tabular}}
\end{table}

Furthermore, we calculate the temperature difference ($\Delta_T$) of both open-office and close-office configurations by calculating the instantaneous temperature difference $\Delta_{T_t} = |T_{t, max} - T_{t, min}|$ between the lowest temperature and the highest temperature of each office, where $t$ represents the time step. 
Then we take the mean value of this parameter over the year. 
Similarly, the average temperature variance ($\sigma^2_T$) is calculated in a similar way where $\sigma^2_{T_t} = var. (T_t)$. Table \ref{tab:heat_trans_delta} shows that the average temperature difference and the average temperature variance of the open offices are smaller than those of the closed office configurations, meaning that the connected space has better heat transferability, as discussed in Section \ref{sec:problem}.

\subsection{Energy Saving Test with DQN}

\begin{figure}[h]
    \centering
    \centerline{\includegraphics[width=1\columnwidth]{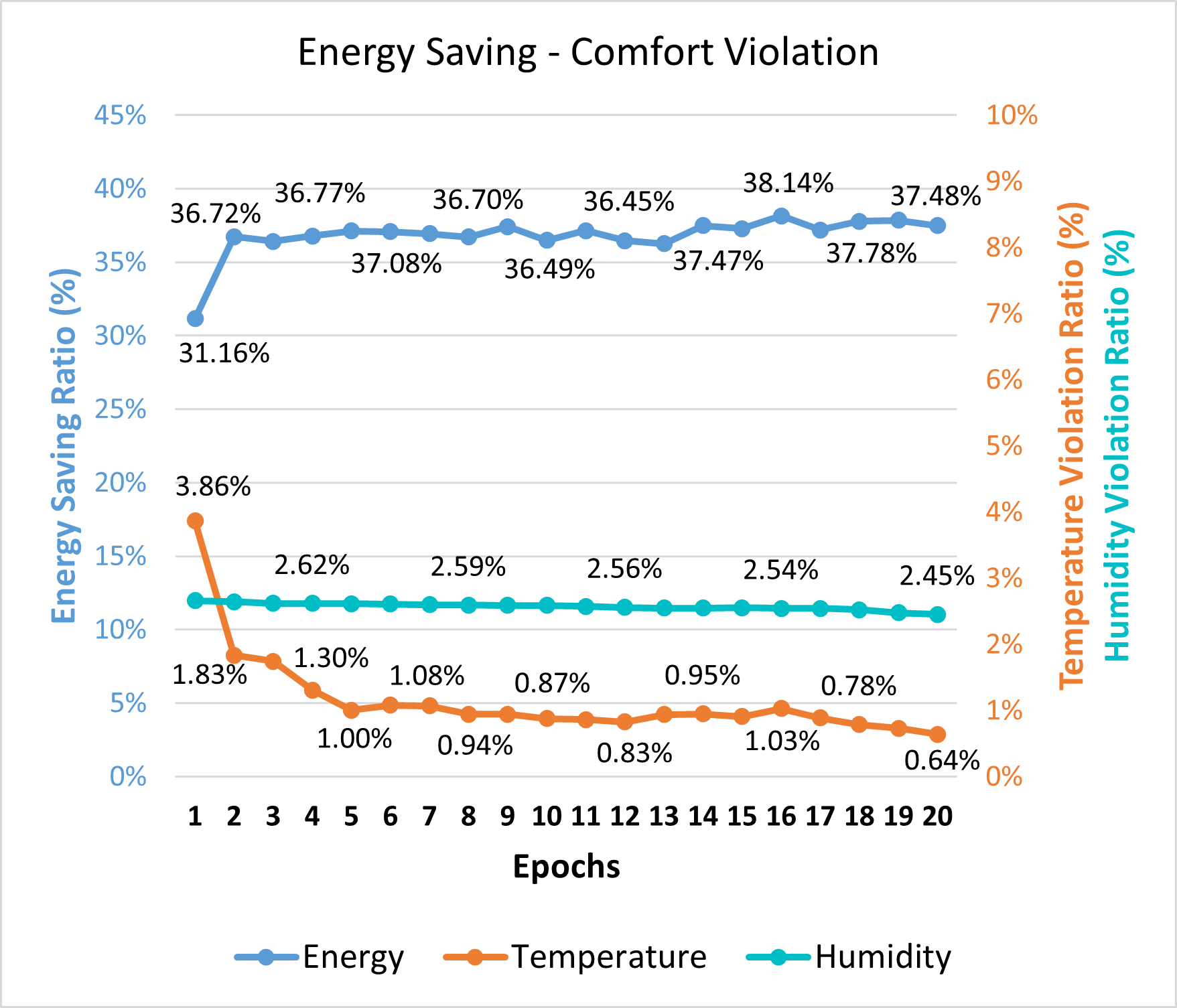}}
    \caption{\hw{HVAC Electricity Consumption and Comfort Violation Over Training Epochs.} }
    \label{fig:E_T}
\end{figure}

Figure \ref{fig:E_T} shows the trends of instantaneous energy saving and comfort violation \ar{for temperature and humidity} during the training process. In short, the energy saving ratio reaches its maximum rapidly as the training starts, while the \ar{temperature} violation rate reduces to under 1\% just at the \hw{fifth} training round. \ar{Also, we observe that the humidity violation declines and remains under 3\% during the entire training phase.} This proves that the proposed method converges fast and can reach an optimum point with a shorter training time than other methods,

To assure that the algorithm has reached its best operation, we let the training continue for 20 epochs, then apply the proposed DQN with the well-trained weights to run the simulation in the test/operation phase. To rapidly deploy the proposed DQN, we usually utilize 5-epoch training as it already reaches an optimum point. The total 5 epochs of training take only 40 minutes, as mentioned in Section \ref{sec:case}, which makes it convenient for practical implementation in real-world buildings.

\begin{figure}[h]
    \centering
    \centerline{\includegraphics[width=1\columnwidth]{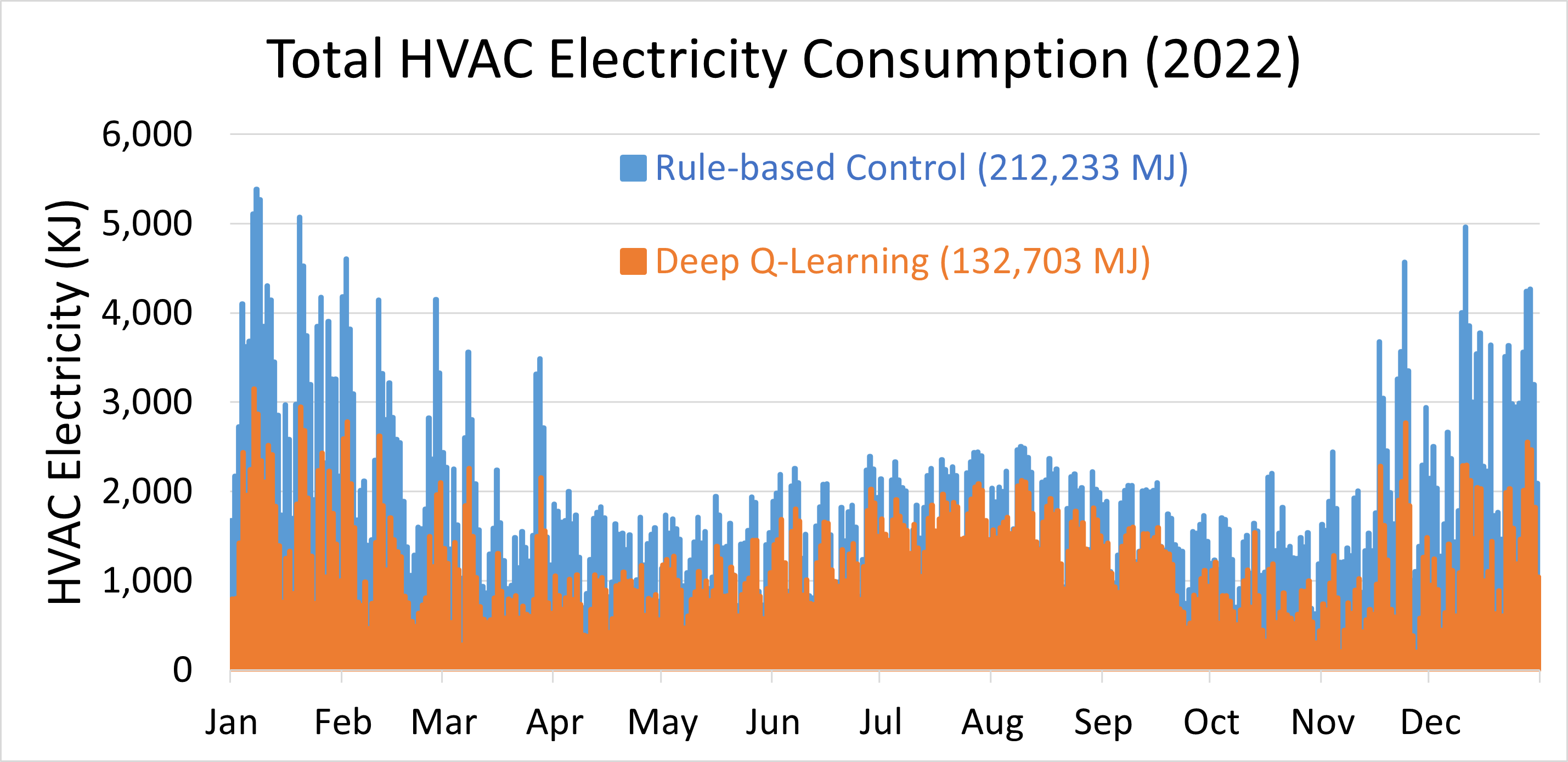}}
    \caption{HVAC Electricity Comparison of the proposed DQN and baseline RBC.}
    \label{fig:E_save}
\end{figure}

The energy consumption when using our DQN model is compared against that of the baseline Rule-based model in Figure \ref{fig:E_save}, where it shows a significant gain in energy efficiency both in the instantaneous and average senses, meaning that the trained DQN can achieve much less energy consumption in total up 37\% with a negligible comfort violation of 1\% and below.

\subsection{Heuristic Reward}

\begin{figure}[h]
    \centering
    \centerline{\includegraphics[width=1\columnwidth]{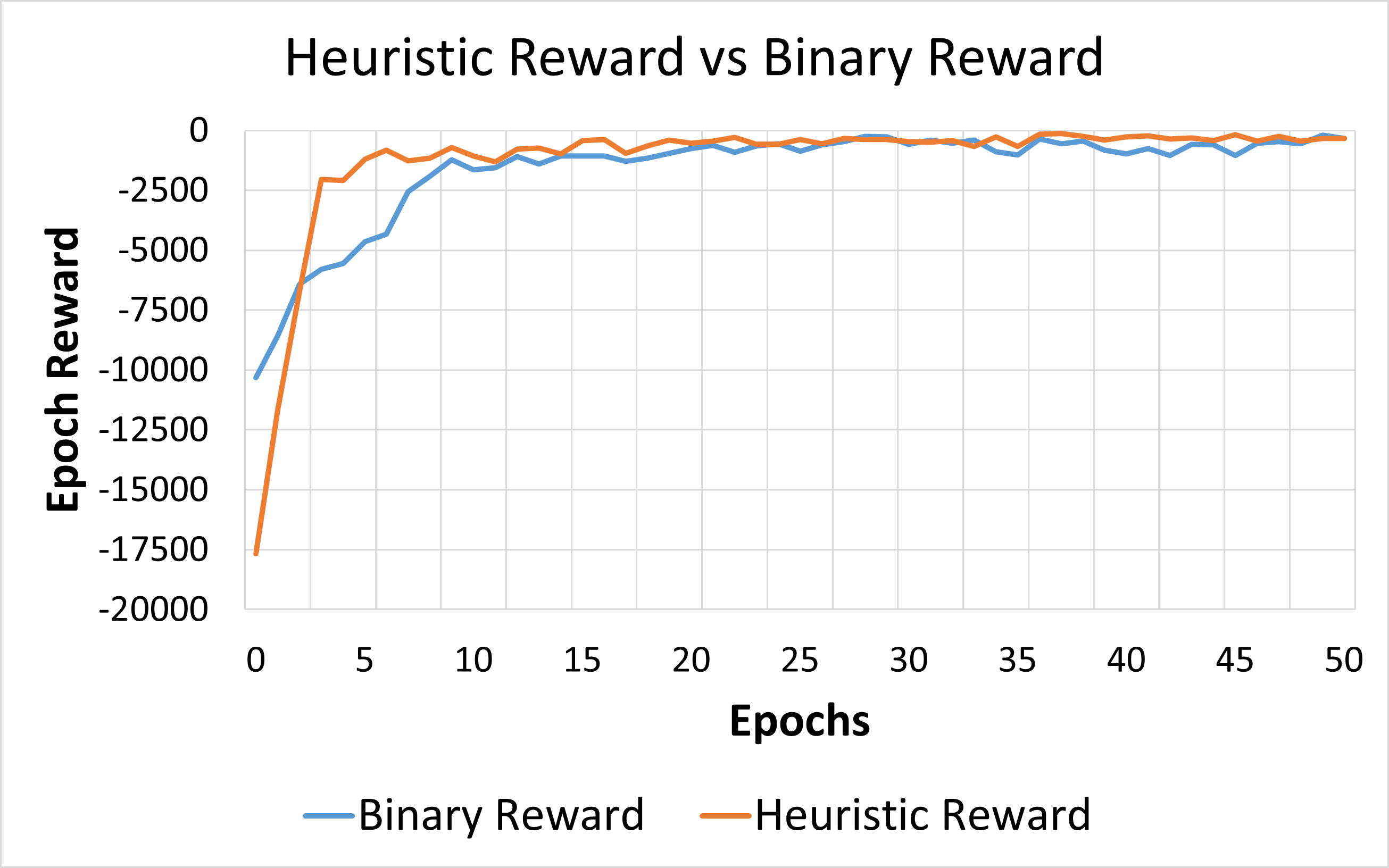}}
    \caption{DQN Training Process in terms of Reward. }
    \label{fig:reward}
\end{figure}

In this study, a heuristic reward mechanism is employed to accelerate the convergence of 
the training process. 
The proposed reward function in Eqs. (\ref{eq:LT}) and (\ref{eq:LTi}) is designed to accurately measure the distance between the desired outcome and the current outcome (temperature), 
as demonstrated in Figure \ref{fig:reward}.

Furthermore, our proposed loss makes a balance between seemingly contradictory objectives of energy efficiency, comfort violation, and smooth operations, as discussed in Section \ref{sec:reward}. 
In detail, Figure \ref{fig:reward} shows that the DQN with our defined heuristic reward explored an optimum point after 5 epochs, while it takes 20 epochs of training of DQN with conventional binary loss to reach the same reward level. On the other hand, our heuristic reward function keeps DQN training stable, while the training process with binary loss still keeps fluctuating even after 40 epochs.

\subsection{Energy-Comfort Trade-Off}
\label{sec:trade}

\begin{figure}[ht]
    \centering
    \centerline{\includegraphics[width=1\columnwidth]{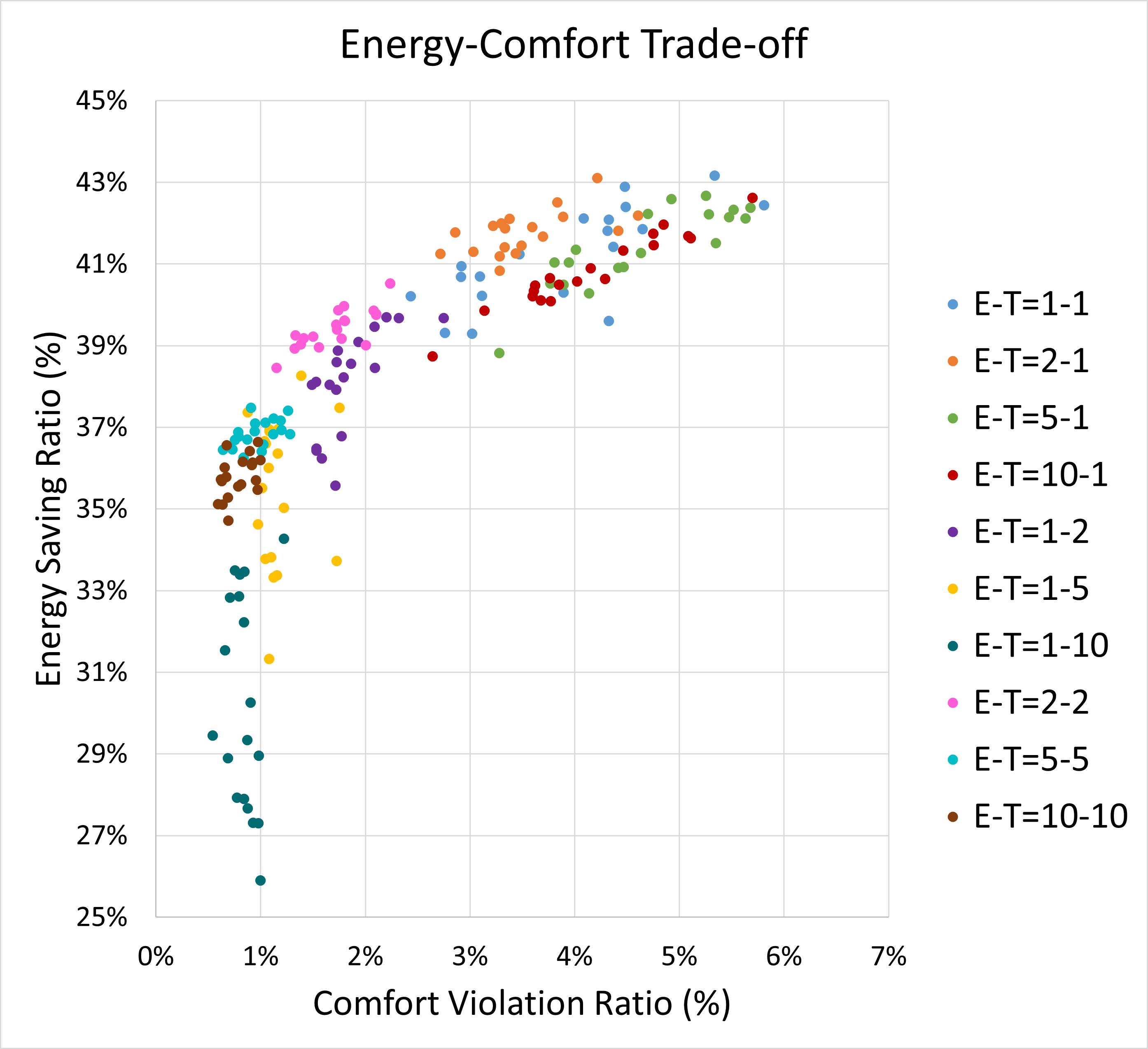}}
    \caption{Trade-off between energy saving and comfort violation. E-T represents the $\eta_E/\eta_T$ ratio}
    \label{fig:E_T_grid}
\end{figure}

We also explore the trade-off between energy efficiency and thermal comfort by fine-tuning the energy savings factor and comfort violation penalties. 
Our results, as depicted in Figure \ref{fig:E_T_grid} and shown on Table \ref{tab:E_T}, indicate that these hyperparameters can significantly influence the model performance.

Specifically, when the energy saving factor is set to $\eta_E=1$, we can observe that by increasing the comfort violation factor from $\eta_T=1$ gradually to $2$, $5$, and $10$, the comfort violation ratio, as well as the energy saving ratio, are obviously decreased. This is because the HVAC system cost more energy to maintain the temperature within a specific range under a more strict comfort policy.

Nevertheless, when the comfort violation factor has been fixed to $\eta_T=1$, the increment of the energy-saving factor did not significantly change the results. 
Interestingly, increasing both factors simultaneously while maintaining the same ratio $\eta_T/\eta_E=1$, can shift the results diagonally, which helps to find a sweet spot that achieves both thermal comfort and energy optimization.


\begin{table}[ht]
\centering
\caption{Energy-Comfort Trade-Off}
\label{tab:E_T}
\resizebox{1\columnwidth}{!}{%
\begin{tabular}{cccc} 
\toprule
E:T Ratio & Electricity (MJ) & Energy Saving (\%) & comfort violation (\%)  \\ 
\midrule
1:1       & 125,040          & 41.08              & 3.86                        \\
2:1       & 123,763          & 41.69              & 3.54                        \\
5:1       & 125,218          & 40.99              & 4.57                        \\
10:1      & 126,399          & 40.44              & 4.10                        \\ 
\hline
1:2       & 131,908          & 37.85              & 1.87                        \\
1:5       & 137,703          & 35.12              & 1.18                        \\
1:10      & 147,819          & 30.35              & 0.86                        \\ 
\hline
2:2       & 128,851          & 39.29              & 1.72                        \\
5:5       & 134,627          & 36.57              & 1.01                        \\
10:10     & 136,778          & 35.55              & 0.94                        \\
\bottomrule
\end{tabular}}
\end{table}

\subsection{Signal Smoothness Trade-Off}
\label{sec:signal}

As discussed in Section \ref{sec:reward}, we proposed a signal consistency penalty to minimize unnecessary VAV control signal fluctuations. 
Specifically, we incorporated the signal loss term $\mathcal{L}_{S}$ to the reward function in (\ref{eq:onoff}) to penalize on/off transitions when yielding small energy efficiency and comfort gains.

\begin{figure}[h]
    \centering
    \centerline{\includegraphics[width=1\columnwidth]{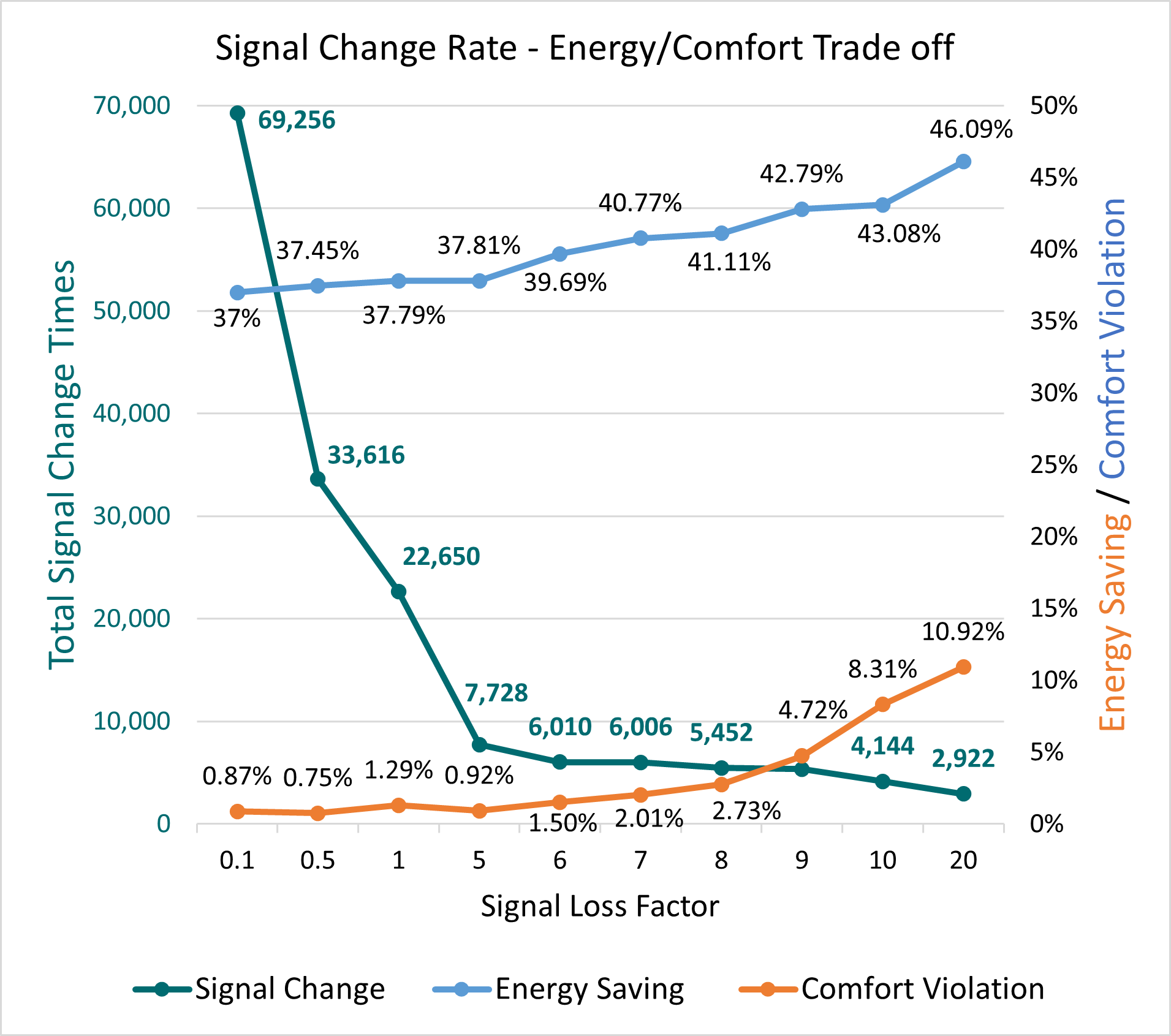}}
    \caption{\hw{The Trade-off between Control Signal's Change Rate and Energy Saving Ratio and Comfort Violation.} }
    \label{fig:signal_2}
\end{figure}

As shown in Figure \ref{fig:signal_2}, the total signal change rate for each VAV unit starts to drop greatly by increasing the weight of smoothness in the objective function by increasing the smoothing tuning factor $\eta_S$. This much smoother operation is achieved without a significant compromise in the comfort level, This indicates that the proposed signal loss term works properly to suppress the frequent control signal variations.

\begin{table*}[h]
\centering
\caption{\hw{Comparative Results for Energy Optimization Using Ours and Alternative Methods.}}
\label{tab:compare}
\resizebox{1\textwidth}{!}{%
\begin{tabular}{cccccccc} 
\toprule
&Energy Saving& Temperature Violation&Humidity Violation& Total Signal Change& Time Cost/epoch (min)   &Time to Convergency& Model Size (Parameters)\\ 
\hline
Baseline&-& 3.68\%&2.42\%& 9,776& 2.09  &-& -\\
PID&28.64\%& 2.30\%&2.94\%& 41,076& 3.18  &-& -\\
MPC&38.16\%& 1.77\%&2.68\%& 46,022& 66.34  &-& -\\
MADDPG&31.15\%& 1.11\%&2.71\%& 30,676& 47.66  &142.98& 17,410 * 12\\
 Ours @ 99\% CCR&37.48\%& 0.64\%&2.45\%& 36,057& 7.75  &38.75& 18,818\\
 Ours @ 98.5\% CCR& 39.22\%& 1.29\%&2.55\%& 6,010& 7.75  &38.75& 18,818\\
 Ours @ 98\% CCR&40.77\%& 2.01\%&2.59\%& 6,006& 7.75  &38.75& 18,818\\
 Ours @ 95\% CCR&42.79\%& 4.72\%&2.66\%& 5,320& 7.75  &38.75& 18,818\\
 
 \bottomrule
\end{tabular}}
\end{table*}

\begin{figure*}[h]
    \centering
    \centerline{\includegraphics[width=1\textwidth]{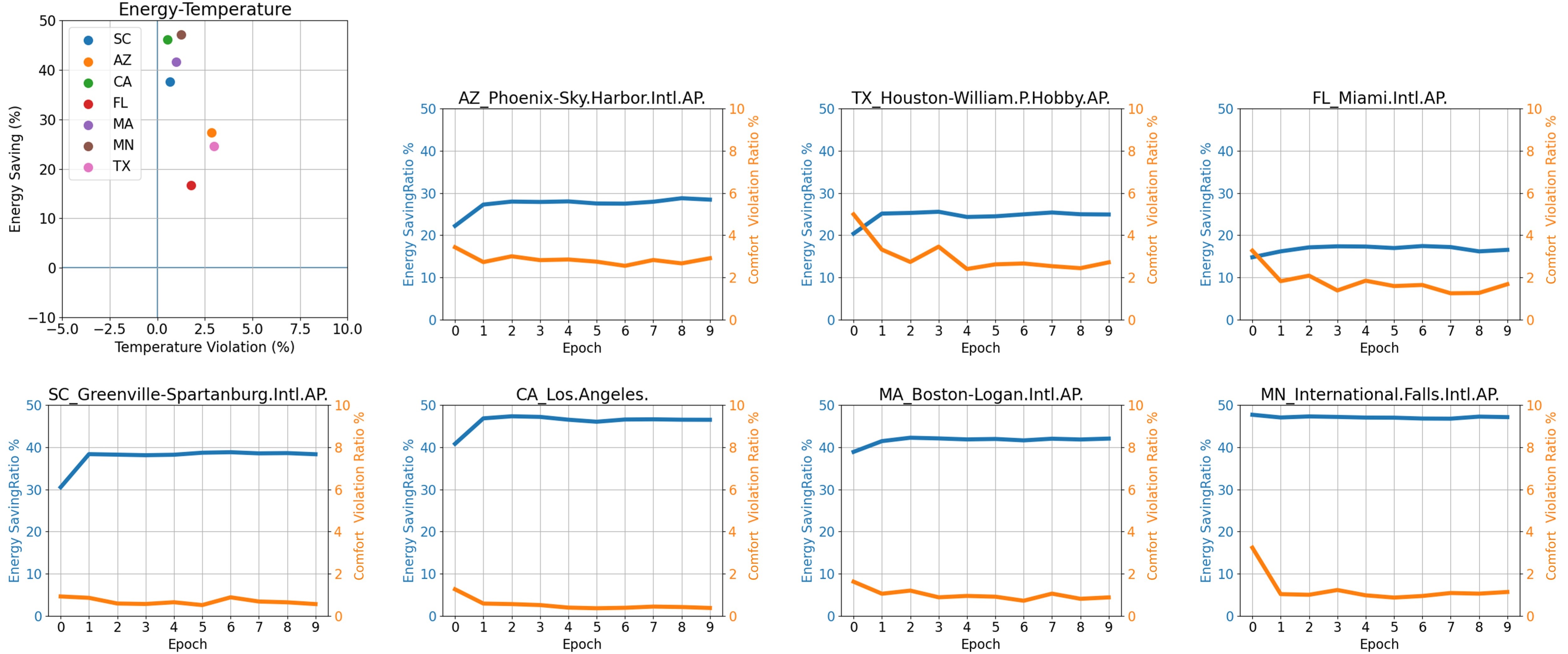}}
    \caption{Test Open Office Model in 7 Different Places.}
    \label{fig:6zone_2_ET}
\end{figure*}

\begin{table*}[h]
\centering
\caption{Test Results for Selected Locations}
\label{tab:weather-data}
\resizebox{1\textwidth}{!}{%
\begin{tabular}{|c|c|c|c|c|c|c|c|c|c|} 
\toprule
Location                & E\_saving (\%) & T\_violation (\%) & T\_offset(F) & T\_var & T\_mean(F) & T\_max(F) & T\_min(F) & Humidity\_mean(\%) & HVAC\_Electricity(MJ)  \\ 
\hline
SC\_Greenville          & 38.83          & 0.51              & 1.80         & 0.80   & 60.17      & 96.08     & 15.98     & 67.81              & 128,683                \\
AZ\_Phoenix             & 28.77          & 2.55              & 1.28         & 0.54   & 74.89      & 111.92    & 35.96     & 34.18              & 119,722                \\
CA\_Los.Angeles         & 47.36          & 0.36              & 1.99         & 0.45   & 62.01      & 95.00     & 39.92     & 69.92              & 69,536                 \\
FL\_Miami               & 17.43          & 1.25              & 0.60         & 0.45   & 76.13      & 96.08     & 41.00     & 72.57              & 159,035                \\
MA\_Boston              & 42.27          & 0.72              & 2.10         & 0.84   & 51.11      & 98.96     & -4.00     & 65.71              & 156,736                \\
MN\_International.Falls & 47.73          & 0.87              & 3.05         & 0.99   & 38.09      & 95.00     & -32.08    & 70.71              & 232,095                \\
TX\_Houston             & 25.56          & 2.40              & 0.99         & 0.53   & 69.97      & 96.98     & 32.90     & 74.27              & 135,768                \\
\bottomrule
\end{tabular}}
\end{table*}

Indeed, Figure \ref{fig:signal_2} shows that the increase of the signal loss factor will also rise both the energy saving ratio and comfort violation. This is because as the total signal change rate reduces, the VAV unit does not change its system state frequently. Consequently, the model tends to simply turn off VAV units to minimize energy losses. Also, we avoid extra energy consumption required for on/off transitions. However, this also contributes to higher comfort violations as the system may not respond swiftly enough to maintain optimal thermal conditions. 
Overall, a sweet point can still be found in this trade-off that keeps both signal change rate and comfort violation low, while maintaining a high energy saving level, when the signal loss factor is between $5$ and $7$. 

\ifx \myver \hugever
No need for this: Additionally, with fine-tuning and a longer training period, the final results still have plenty of room to improve. 
\fi

\hw{
\subsection{Comparison with Other Methods}
}

\hw{
In this section, we compare our model with the most commonly used conventional control methods, such as PID \cite{homod2009pid} and MPC \cite {yao2021state}. We also compare it with MADDPG (Multi-Agent Deep Deterministic Policy Gradient), an advanced deep reinforcement learning model for complex tasks \cite{yu2020multi}.
}

\hw{
To ensure the equality of comparison, we applied exactly equal building layout and simulation settings, provided in Section \ref{sec:office_model}, to all methods.
As shown in Table \ref{tab:compare}, the results indicate that our method outperforms other approaches from different perspectives. Compared to PID, our method achieves a substantially higher energy efficiency. Our method saves between 37\% in energy consumption (with respect to baseline), while PID archives only 28\% energy saved. This enhanced energy efficiency comes at a lower comfort violation for our model, meaning that our method fully dominates the PID method. 
Compared to MPC, our model is more time efficient and can reach a higher energy-saving rate by trading the comfort violation. Compared to MADDPG, our method achieves a higher energy efficiency (37\% compared to 31\%) while consuming less time to converge (9-fold improvement). Ours is also more computationally efficient than MADDPG, meaning that it is more suitable to be deployed in building energy management systems that has limited performance and requires real-time calculation. 
}

\hw{
Furthermore, our method has the additional advantage of minimizing signal change frequency to prolong the lifetime of HVAC system's mechanical components. It also reduces the occupants' discomfort by avoiding unnecessary transitions. Finally, it enhances energy efficiency since typical HVAC systems consume higher energy levels during transition intervals.
}

\subsection{Model Generalizability Test}
\label{sec:general}

\begin{figure}[h]
    \centering
    \centerline{\includegraphics[width=1\columnwidth]{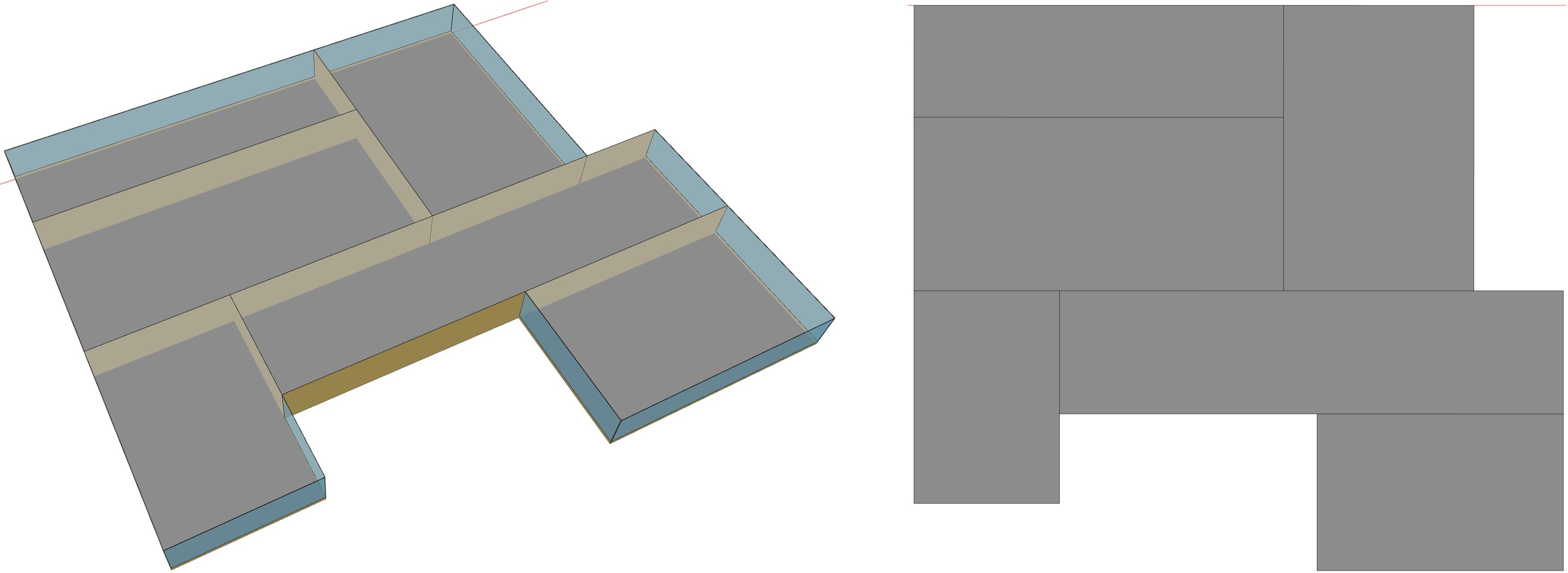}}
    \caption{A Test Open Office Model with 6 sub-zones. }
    \label{fig:6zone_2}
\end{figure}

Here, we examine the applicability of the proposed deep reinforcement learning framework by applying it to a new open office with a different floor plan and a different thermal zone partition and evaluating the achieved energy efficiency and occupant comfort. 
The new open office design is also comprised of six thermal zones, as depicted in Figure \ref{fig:6zone_2}.

\begin{figure*}[h]
    \centering
    \centerline{\includegraphics[width=1\textwidth]{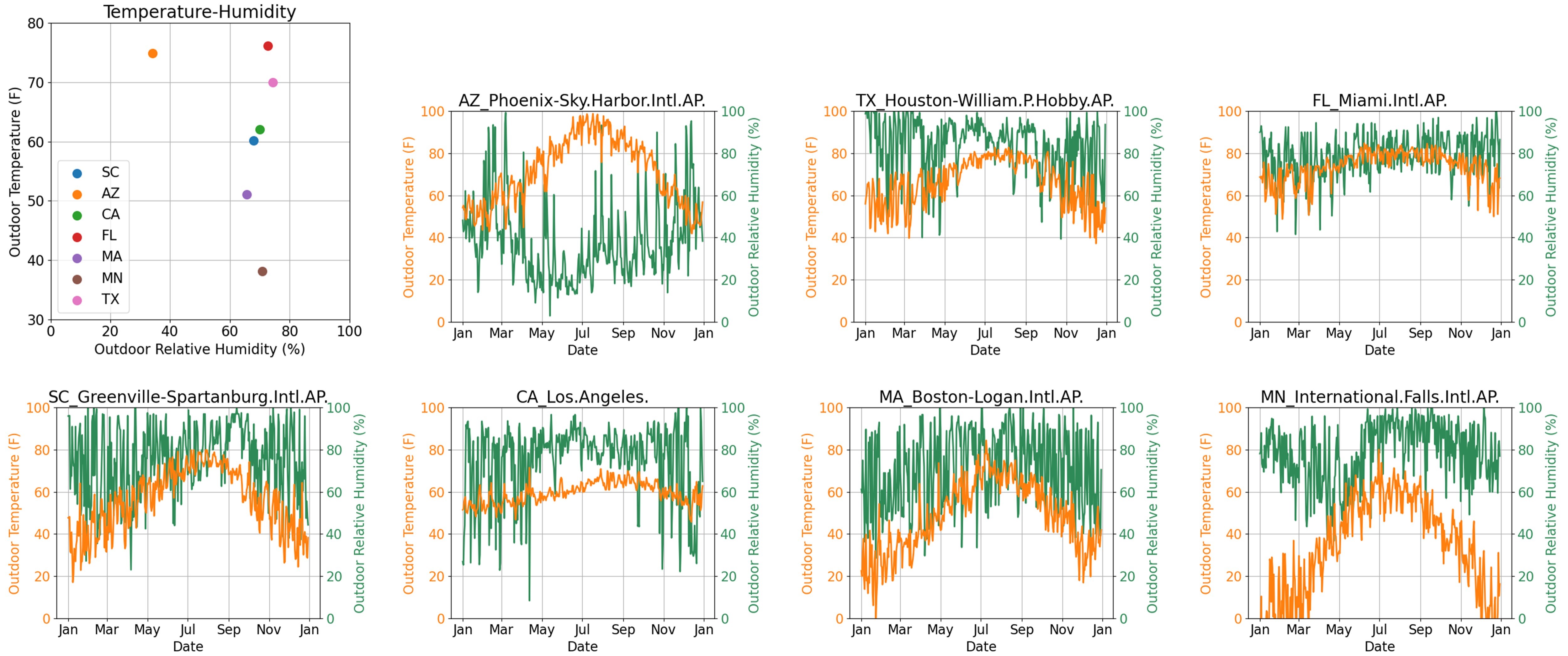}}
    \caption{Weather Condition in 7 Different Places.}
    \label{fig:6zone_2_weather}
\end{figure*}

\ifx \myver \hugever
Sounds duplicate:
Through the implementation of the proposed DQN method, the energy consumption of the new office model will be optimized while maintaining thermal comfort for the occupants. The results of this study will further demonstrate the flexibility and versatility of the proposed DRL framework in optimizing energy consumption in open-plan offices.
\fi
The results of this test are presented in Figure \ref{fig:6zone_2_ET}, which demonstrate its efficacy in reducing energy consumption across a range of weather conditions. The method is therefore applicable to different weather locations, including hot regions such as Arizona, Texas, and Florida, as well as cold regions such as Minnesota and Massachusetts. 

We also observe that the performance remains high under varying humidity levels as well.
The results show that the proposed method can achieve substantial energy savings, up to 40\%, in regions 
such as Los Angeles, Boston, and International Falls. However, in regions with more extreme weather conditions, such as Phoenix, Houston, and Miami, energy savings are limited from 17\% to 30\% due to the unique weather conditions of these areas. These findings also highlight the importance of considering specific weather conditions on the expected energy efficiency when optimizing energy consumption in open-plan offices.

The weather data of the above seven different locations are presented in Figure \ref{fig:6zone_2_weather}, highlighting the interplay between the weather conditions and energy optimization. This analysis reveals that regions with higher mean outdoor temperatures (above 70°F) present a greater challenge in reducing energy consumption, as indicated by the higher levels of comfort violation in these areas. Conversely, regions with lower temperatures offer more opportunities for energy savings. This finding highlights the importance of considering the unique weather conditions when optimizing energy consumption in open-plan offices.

Table \ref{tab:weather-data} shows the equivalent numerical results of the test over a year, where \textit{E\_saving} is the energy saving ratio, \textit{T\_violation} is the comfort violation ratio, \textit{T\_offset} is the offset of the comfort violation degree, \textit{T\_var} is the variance of the zone temperatures among a year, \textit{T\_mean} is the average zone temperatures among a year, \textit{T\_max}, \textit{T\_min}, and \textit{Humidity\_mean} are the maximum temperature, the lowest temperature, and the average humidity of the local city, and \textit{HVAC\_Electricity} is the total HVAC electricity cost in a year.

\section{\ar{Discussions}}

\ar{It is noteworthy that the number of actions in our proposed DRL framework grows exponentially with the number of vents, which may require extremely long episodes for the network to converge (even after using experience replay). In our case, we had only 6 VAV boxes in the open office, which resulted in a total number of $2^6 = 64$ actions. However, once we increase the number of VAVs in an open office to 20, the number of total actions becomes $2^{20}=1,048,576$. Incorporating some heuristic optimization methods such as genetics algorithm, and PSO into the framework can alleviate this issue. 
}

\hw{
Also, capturing body heat through occupancy rate always includes an inherent approximation, and including computer vision methods to have a better count of occupants can enhance the performance of the method. This work can also be extended not only to control the on-off policy of existing vents, but also to offer design solutions by moving the positions of the vents and adding/removing new vents to the system.}

\section{Conclusion}
\label{sec:conclusion}

The increasing demand for convenient communication and improved energy efficiency has led to the popularity of open offices with multi-VAV HVAC systems as a global trend. However, optimizing energy usage in such environments requires a careful balance between thermal comfort, health considerations, and energy efficiency, particularly in the post-COVID era, where some building zones have reduced working hours or fewer occupants due to remote working policies. 

However, the majority of AI-based energy optimization methods that are primarily developed for close plans, unfortunately, do not suit open plans for a number of issues such as being over-complicated and lack of generalizability, requiring prohibitively long training time, ignoring the inter-zone heat transfer in zone-level optimization, and manipulating factors that are not easily accessible. 
This paper addresses this research gap in Deep Reinforcement Learning (DRL)-based HVAC control, specifically tailored for open offices. Our approach not only overcomes the above-mentioned challenges faced by existing methods but also provides a general solution that is both flexible and simple to implement. Training takes only a few minutes compared to weeks and months of competitor methods. In addition, we conducted a comprehensive analysis of the thermodynamic characteristics of open offices and identified the specific energy management strategies required. By applying our proposed method to accumulated data from a sample building, we achieved a 37\% reduction in HVAC energy consumption compared to the baseline rule-based method while maintaining comfort violations below 1\%. Furthermore, we introduced a novel signal smoothness control term to prevent mechanical wear and occupant discomfort due to unnecessary frequent on/off transitions. 
Finally, we showed the generalizability of our model by applying it to different plans under substantially different weather conditions, which yielded high performance in diverse environments.

\section*{Acknowledgments}

The authors would like to thank Dr. M.Z. Naser for his comments on HVAC system design principles and thermodynamic theories. This research is partially supported by BMW Manufacturing Co. LLC under grant \#4500646308.

 \bibliographystyle{elsarticle-num} 
 \bibliography{cas-refs}





\end{document}